\documentclass[aps,pra,twocolumn,notitlepage,showpacs,superscriptaddress,nofootinbib]{revtex4-1}
\usepackage{graphicx}
\usepackage{bm}
\usepackage{amsmath}
\usepackage[usenames,dvipsnames]{color}
\usepackage{amsfonts}
\usepackage[colorlinks]{hyperref}
\usepackage[english]{babel}
\usepackage[normalem]{ulem}
\usepackage{overpic}
\usepackage{mathptmx}

\newcommand{\br}{{\bf r}}

\newcommand{\bp}{{\bf p}}

\newcommand{\ee}{\end{equation}}
\newcommand{\be}{\begin{equation}}

\newcommand {\bq} {{\boldsymbol {q}}}

\newcommand {\Be} {\begin{eqnarray*}}
\newcommand {\Ee} {\end{eqnarray*}}
\newcommand {\beq} {\begin{eqnarray}}
\newcommand {\eeq} {\end{eqnarray}}
\newcommand {\Bi} {\begin{enumerate} }
\newcommand {\Ei} {\end{enumerate}}
\newcommand {\Bs} {\left \{ \begin{array}{l} }
\newcommand {\Es} {\end{array} \right . }
\newcommand {\Bss} {\left \{ \begin{array}{ll@{\quad}l}}
\newcommand {\Ess} {\end{array} \right . }
\newcommand{\eps}{\epsilon}
\definecolor{darkgreen}{rgb}{0,0.6,0}

\date{\vspace{-5ex}}

\newcommand\redout{\bgroup\markoverwith
{\textcolor{red}{\rule[0.5ex]{2pt}{0.8pt}}}\ULon}

\begin{document}

\title{Quasi-thermalization of collisionless particles in quadrupole potentials}
\author{Johnathan Lau \textdagger}
\affiliation{Mathematical Sciences, University of Southampton, Highfield, Southampton, SO17 1BJ, United Kingdom}
\footnote{\textdagger Deceased 21 May 2016}
\author{Olga Goulko}
\affiliation{Department of Physics, Boise State University, Boise, ID 83725, USA}
\author{Thomas Reimann}
\affiliation{Laboratoire Kastler Brossel, ENS-Universit\'e PSL, CNRS, Sorbonne Universit\'e, Coll\`ege de France, 24 rue Lhomond, 75005 Paris, France}
\author{Daniel Suchet}
\affiliation{Laboratoire Kastler Brossel, ENS-Universit\'e PSL, CNRS, Sorbonne Universit\'e, Coll\`ege de France, 24 rue Lhomond, 75005 Paris, France}
\affiliation{Institut du Photovolta\"ique d'\^Ile de France (IPVF), UMR 9006, \'Ecole polytechnique, IP Paris, Palaiseau, France}
\author{C\'edric Enesa}
\affiliation{Laboratoire Kastler Brossel, ENS-Universit\'e PSL, CNRS, Sorbonne Universit\'e, Coll\`ege de France, 24 rue Lhomond, 75005 Paris, France}
\author{Fr\'ed\'eric Chevy}
\affiliation{Laboratoire Kastler Brossel, ENS-Universit\'e PSL, CNRS, Sorbonne Universit\'e, Coll\`ege de France, 24 rue Lhomond, 75005 Paris, France}
\author{Carlos Lobo}
\affiliation{Mathematical Sciences, University of Southampton, Highfield, Southampton, SO17 1BJ, United Kingdom}
\date{\today}

\begin{abstract}
We analyze several puzzling features of a recent experiment with a noninteracting gas of atoms in a quadrupole trap. After an initial momentum kick, the system reaches a stationary, quasi-thermal state even without collisions, due to the dephasing of individual particle trajectories. Surprisingly, the momentum distribution remains anisotropic at long times, characterized by different ``temperatures" along the different directions. In particular, there is no transfer of the kick energy between the axial and radial trap directions. To understand these effects we discuss and solve two closely related models: a spherically symmetric trap $V(r)\simeq r^\alpha$ and a strongly confined gas along one direction (a ``pancake" trap). We find that in the isotropic trap, the gas unexpectedly also preserves the anisotropy of the kick at long times, which we are able to explain using the conservation of angular momentum and the virial theorem. Depending on the value of $\alpha$ we find that the kick can cool or heat the orthogonal directions. The pancake trap case is quantitatively similar to the quadrupole one. We show that for the former, the temperature anisotropy and memory of the kick direction are due to the change in the 2D effective potential resulting from the kick, thereby also explaining the quadrupole experimental results.
\end{abstract}


\maketitle

\section{Introduction}
\label{sec:Intro}

A major part of the theoretical study of classical Hamiltonian dynamics  \cite{Arnold2013} concerns the ability of purely conservative systems to reach thermal equilibrium. 
This line of inquiry took its origin from Boltzmann's demonstration of the celebrated H-theorem, which provided for the first time a microscopic explanation of the second law of thermodynamics. This pioneering work quickly gave rise to many paradoxes due to the fact that the entropy of a Hamiltonian system $\cal S$  is conserved. Even 150 years after Boltzmann's work, problems like Loschmidt's paradox \cite{loschmidt1876uber}, Poincar\'e's recurrence theorem \cite{poincare1890probleme} or the Fermi-Pasta-Ulam-Tsingou \cite{fermi1955studies,dauxois2008fermi} problem remain largely unresolved  \cite{savitt1995time}. A possible approach towards their resolution lies in the notion that any small sub-ensemble ${\cal S}_1$ can be described as an open system interacting with the rest of $\cal S$. The latter therefore plays the role of a bath allowing the thermalization of ${\cal S}_1$. In the quantum world, this picture is known as the Eigenstate Thermalization Hypothesis \cite{deutsch91quantum,sredniki94chaos,rigol2008thermalization}, which also applies to classical systems \cite{jin2013equilibration}. In this context, an intriguing question concerns the minimal system size required for thermalization. Despite the fact that the thermodynamic limit is usually associated with large-size systems, small objects such as nanoparticles \cite{voisin2000size}, nuclei \cite{sarkar2017thermalization} or atoms trapped in optical lattices \cite{kaufman2016quantum} are nevertheless known to relax towards thermal equilibrium.

With decreasing system size, a natural question to consider is to which degree a single particle is able to reach thermal equilibrium. This extreme limit can be studied experimentally using cold fermions by taking advantage of the Pauli exclusion principle. The associated suppression of interactions at low temperature gives rise to a unique experimental platform facilitating the study of purely Hamiltonian systems.

In this work, we consider an ensemble of non-interacting particles confined in non-separable power law potentials. The question of thermalization in this class of potentials was already raised in the context of collisionless atoms in quadrupole traps \cite{Surkov1994,Davis1995}. This problem was recently revived in the context of quantum simulation of high-energy physics, where the behavior of (harmonically confined) massless Weyl fermions was studied experimentally using cold atoms in a quadrupole trap \cite{Suchet2015}. In this latter work, it was shown that after a rapid quench of the trap position, the center of mass motion is damped after a few oscillations and the system reaches a steady state characterized by partial thermalization of its momentum degrees of freedom. The corresponding distribution of the atomic ensemble closely resembles a thermal distribution, $np_{i=x,y,z}\propto \exp(-p_i^2/2mk_B T_i)$, but with anisotropic temperatures. 

In this paper, we present a detailed theoretical analysis of these relaxation dynamics. Furthermore, we provide analytical calculations of the steady state properties in an isotropic as well as in a pancake geometry. These results are compared to numerical solutions of the corresponding dynamical equations. Our work clarifies the memory effect leading to the anisotropy of the momentum distribution and predicts a singular behavior for spherical potentials.

\section{Relaxation dynamics in quadrupole traps}\label{sec:simulations}


Motivated by recent experiments with non-degenerate spin-polarized fermions \cite{Suchet2015}, we consider an ensemble of classical noninteracting particles confined by a quadrupole trapping potential 
\begin{equation}
V(\mathbf{r}) = \mu_B b \sqrt{x^2 + y^2 + 4 z^2}. \label{eq:potential}
\end{equation}
where $\mu_B$ is the Bohr magneton and $b$ is the magnetic field gradient, a positive quantity.

This potential is {\it non-integrable} since it has three degrees of freedom but only two constants of the motion (total energy $E$ and angular momentum $L_z$). As a consequence its dynamics exhibits chaotic behaviour in some regimes. In contrast, the more usual potential of standard atomic traps is a sum of harmonic terms of the form $V_1(x)+V_2(y)+V_3(z)$ allowing us to define three conserved energies, leading to an integrable problem. Note that, since the quadrupole potential cannot be written as the sum of potentials as in the harmonic case, the motion along one direction depends on the other two so that momentum and energy are constantly being exchanged between the three directions as the atom moves along the orbit.

We will study the {\em relaxation dynamics} in this potential i.e., what happens to the gas after it is slightly perturbed from equilibrium. At $t=0$ with the gas in thermal equilibrium, the atoms receive a ``momentum kick'' $\bq$ that shifts every atom's momentum $\bp \rightarrow \bp+ \bq$ and increases its energy by $\bp \cdot \bq/m +q^2/2m$. Since the original (thermal) distribution before the kick is an even function of each component of $\bp$, the first term drops out when averaged over that distribution, so that the average energy change $\Delta E$ per atom is:
\begin{equation}
\Delta E =q^2/2m. \label{eq:deltaE}
\end{equation}
We are interested in the subsequent evolution: how the gas relaxes to steady state and how the energy $\Delta E$ of the kick is redistributed along the different directions of motion. Normally, as is usually assumed, collisions would be responsible for this redistribution leading to a return to thermal equilibrium. However, in our case, there are no collisions nor mean-field interactions, so any relaxation process is due purely to the nonintegrability of the potential.

The state of the gas can be described by the Boltzmann distribution $f(\mathbf{r},\mathbf{p},t)$ which we normalize to unity:
\begin{equation}
 \int d^3 \mathbf{r} \int d^3 \mathbf{p} \  f(\mathbf{r},\mathbf{p},t)=1.
\end{equation}
All extensive quantities are to be taken as ensemble averages over this distribution. For example, the final measured momentum distribution $np_z$ is given by
\begin{equation}
 np_z= \int d^3 \mathbf{r} \int dp_x \ dp_y \  f(\mathbf{r},\mathbf{p},t \rightarrow \infty). \label{eq:doublyintegrated}
 \end{equation}
 
\begin{figure}

	\begin{overpic}[width=0.9\linewidth]{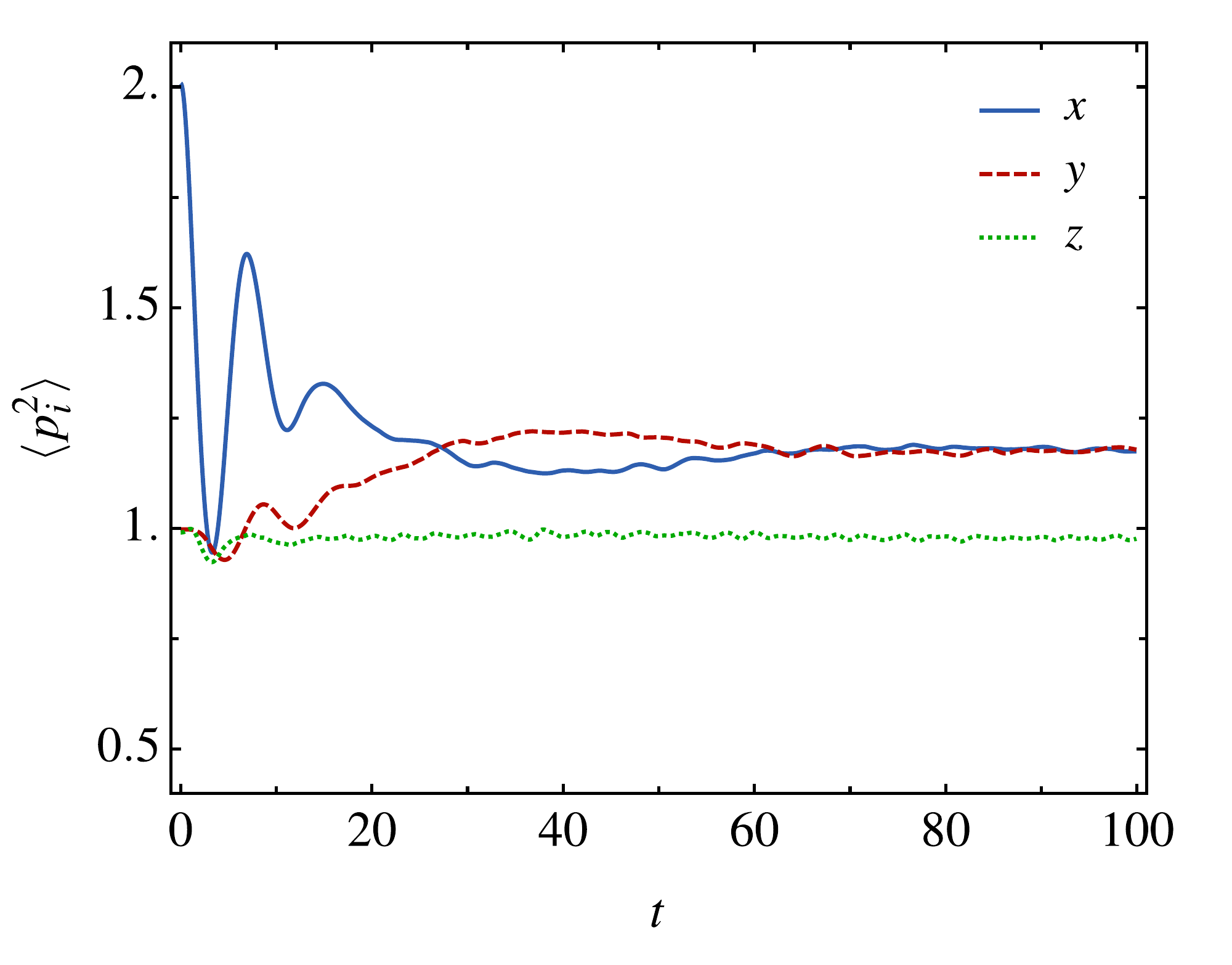}
		\put(1,73){\rm\bfseries a)}
	\end{overpic}
	
	\begin{overpic}[width=0.9\linewidth]{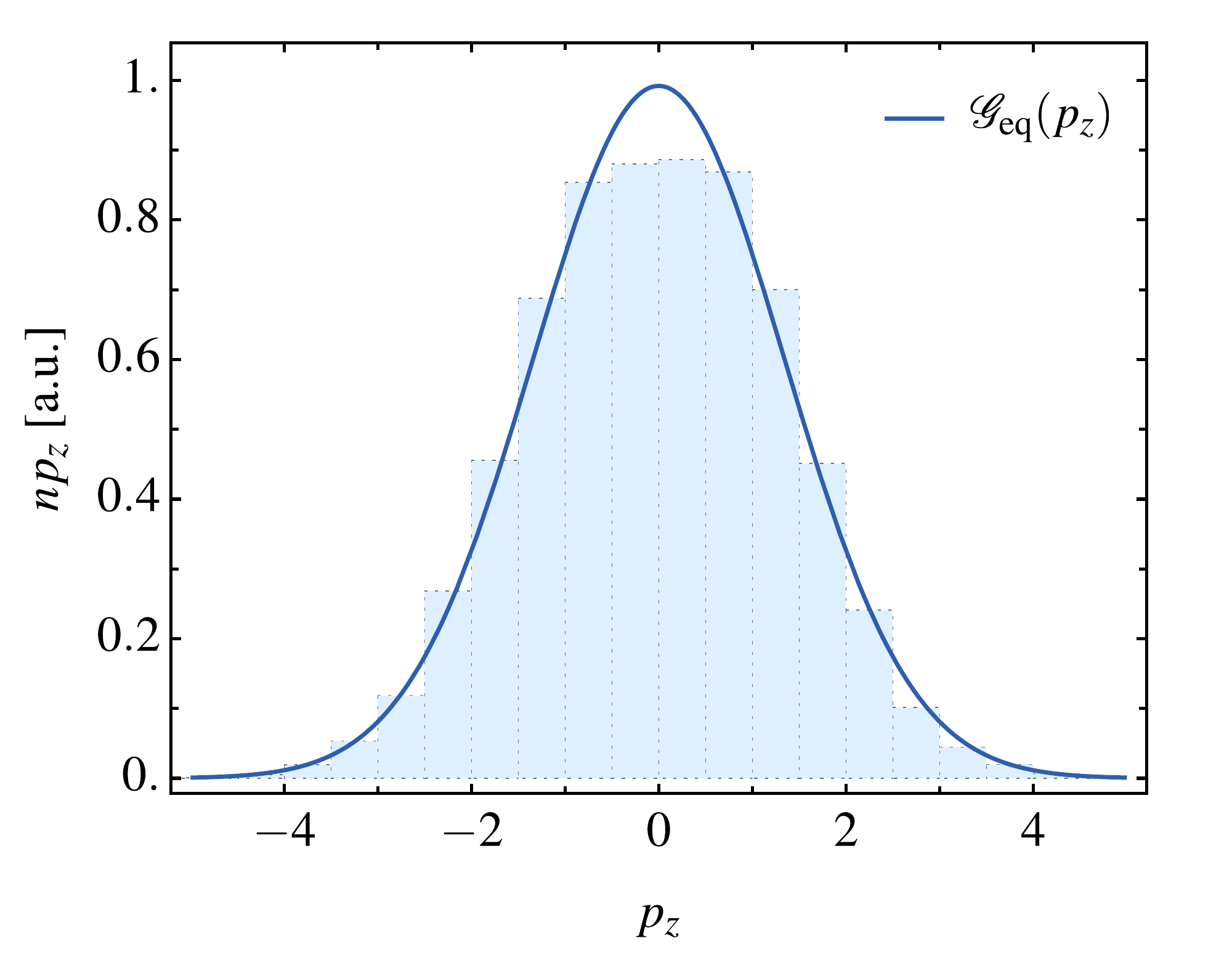}
		\put(1,73){\rm\bfseries b)}
	\end{overpic}
	
	\caption{Numerical simulation of the relaxation dynamics in a quadrupole trap. a) Kinetic energy per atom along different directions as a function of time after a kick $q=1$ along $x$ at $t=0$. The energies seem to reach a stationary state for $t\geq 80$. Along $z$ the average kinetic energy is almost unchanged from its initial value $\sim 1$, but along $x$ and $y$ the corresponding values increase by the same amount to a final value of $\sim 1.2$. b) Momentum distribution $np_z$ of the steady state. The solid line $\mathcal{G}_{\rm eq}$ is a Gaussian distribution with the same variance.\label{fig:simul1}}
\end{figure}
 
To simulate this distribution we perform molecular dynamics simulations of the gas \cite{boltzmannletter, goulko2012boltzmann} where the trajectory of each atom is calculated following the classical equations of motion, without suffering any collision. This method gives us full access to all observables, including the Boltzmann distribution itself. For example, we can measure the phase-space average $\left< p_i^{2} \right>$ for $i=x,y,z$ over the entire system as a function of time by averaging over the trajectories of individual atoms:
\begin{equation}
 \left< p_i^2 \right>_t \equiv \int d^3 \mathbf{r} \int d^3 \mathbf{p} \ p_i^2 f(\mathbf{r},\mathbf{p},t) \simeq \frac{1}{N}\sum_{\mbox{all $N$ atoms}} p_i^2(t), \label{eq:phase_space_average}
\end{equation}

We start with a gas of $N=10^5$ atoms sampled from the initial Boltzmann distribution at temperature $k_B T_0=1$ with a momentum kick of $q$ along $x$:
\begin{equation}
f \propto \exp \left(-\frac{(p_x-q_x)^2+p_y^2+p_z^2}{2} -V(x,y,z)\right) \label{eq:xkickeddistribution}
\end{equation}
(analogously for a kick $q_z$ along $z$ etc.) and let each individual atom evolve according to the classical trajectory. From now on we set $m=k_BT_0=\mu_Bb=1$, which is equivalent to choosing $m$ as the mass unit, $l_0=k_BT_0/\mu_Bb$ as the unit of length and $t_0=\sqrt{m k_BT_0}/\mu_Bb$ as the unit of time. The time evolution is calculated using the velocity Verlet algorithm \cite{Verlet, urban}. We use a time step $\Delta t=0.001t_0$, which provides sufficient accuracy, as the error of the algorithm is of the order $\mathcal{O}(\Delta t^2)$.

This very simple setup gives rise to some surprises which have also been confirmed experimentally \cite{Suchet2015}:
\begin{enumerate}
\item {\em Stationary ``thermal" distribution}: In Fig.~\ref{fig:simul1} a) we plot $\left< p_i^2 \right>_t$. We see that, at long times, it has reached an apparently stationary distribution. In Fig.~\ref{fig:simul1} b) we plot the long time doubly integrated momentum distribution $np_z$ (\ref{eq:doublyintegrated}) and we see that it fits closely to a Gaussian (thermal) distribution $np_z\propto \exp (-p_z^2/2mk_BT_z)$ where we define an effective temperature analogously to the experiment \cite{Suchet2015}:
\begin{equation}
T_i \equiv  \left< p_i^2 \right>_{t \rightarrow \infty}, \mbox{ } i=x,y,z
\end{equation}
so that
\begin{equation}
\Delta T_i \equiv \left< p_i^2 \right>_{t \rightarrow \infty} -\left< p_i^2 \right>_{t =0}.
\end{equation}

\item {\em Anisotropic temperatures}: From Fig.~\ref{fig:simul1} a) we see that, even though the doubly integrated distributions $np_i$ along the different directions $i$ are Gaussian, their widths are different: generally we find $T_x \sim T_y \neq T_z$. We also see that $T_z \sim T_0 < T_{x,y}$ which we find to be true whenever the kick is in the $xy$ plane, the opposite being true if the kick is along the $z$ direction. This is unexpected because the quadrupole potential is non-separable, continuously transfering energy and momentum between all directions for each atom, so we might expect na\"ively that on average $T_x \sim T_y \sim T_z$, i.e.\ there would be a certain degree of ergodicity.
	
\item {\em Apparent separability of the $z$ and $x-y$ distributions}: for a kick along $z$, the width of the momentum distribution along $x$ and $y$ seems to be unchanged (i.e. $T_{x,y} \simeq 1$)  whereas $T_z$ increases. The energy increase $\Delta E$ due to the kick is mainly concentrated into the $z$ direction so that $\Delta E\simeq 3/2 \mbox{ } \Delta T_z$. Likewise, if the kick is along $x$, the increase in kinetic energy along the $z$ is negligible ($T_z\simeq 1$) but {\it both} $T_x$ and $T_y$ increase by the same amount ($T_x=T_y$) so that $\Delta E \simeq 3 \Delta T_x$. In fact we will see below that this separation is not exact; there is a slight increase of energy in directions transverse to the kick. Nevertheless this behaviour is consistent with a strong separation of the dynamics into $z$ and $xy$ plane components even though the potential is non-separable.
\end{enumerate}

The na\"ive, straightforward conclusion from these observations is that {\it the gas seems to have thermalized in the absence of collisions (since the doubly-integrated momentum distributions  \eqref{eq:doublyintegrated} become Gaussian-like, a hallmark of thermalization) but with some effective ``decoupling" of the motion along $z$ and $xy$ directions leading to different temperatures $T_z$ and $T_{xy}$.}

\subsection{Apparent Thermalization} \label{sec:thermalisation}
In point 1. above we noted that the gas becomes stationary after some time. This stationary state of the gas is not due to collisions but to the fact that, in the quadrupole trap, the orbits of different atoms will have different, incommensurate periods leading to the relative dephasing of individual trajectories. This dephasing, when averaged over the whole gas, leads to a stationary distribution. Note that the appearance of a stationary distribution would not happen in the standard harmonic trap since a momentum kick would lead to undamped oscillations of the center of mass. Note also that that irreversibility has not set in by this stage since there are no collisions.

We also mentioned above that the gas seemed to have thermalized in the absence of interactions since the doubly-integrated momentum distributions \eqref{eq:doublyintegrated} become Gaussian after the kick.

Of course, since the effective temperatures deduced from the width of the Gaussians are different ($T_z \neq T_{x,y}$) the state cannot be a true thermal state. Indeed, collisions are necessary to redistribute the kick energy $\Delta E$ among all accessible phase space regions of energy $E+\Delta E$ so that the entropy increases $S(E) \rightarrow S(E+\Delta E)$ whereas in this experiment, $E\rightarrow E+\Delta E$ but entropy is unchanged. Nevertheless, as can be seen from the simulations, several ``thermal" properties {\it can} be achieved, e.g. stationarity and equilibration of temperatures along the $x$ and $y$ directions.

\begin{figure}
\includegraphics[width=0.9\linewidth]{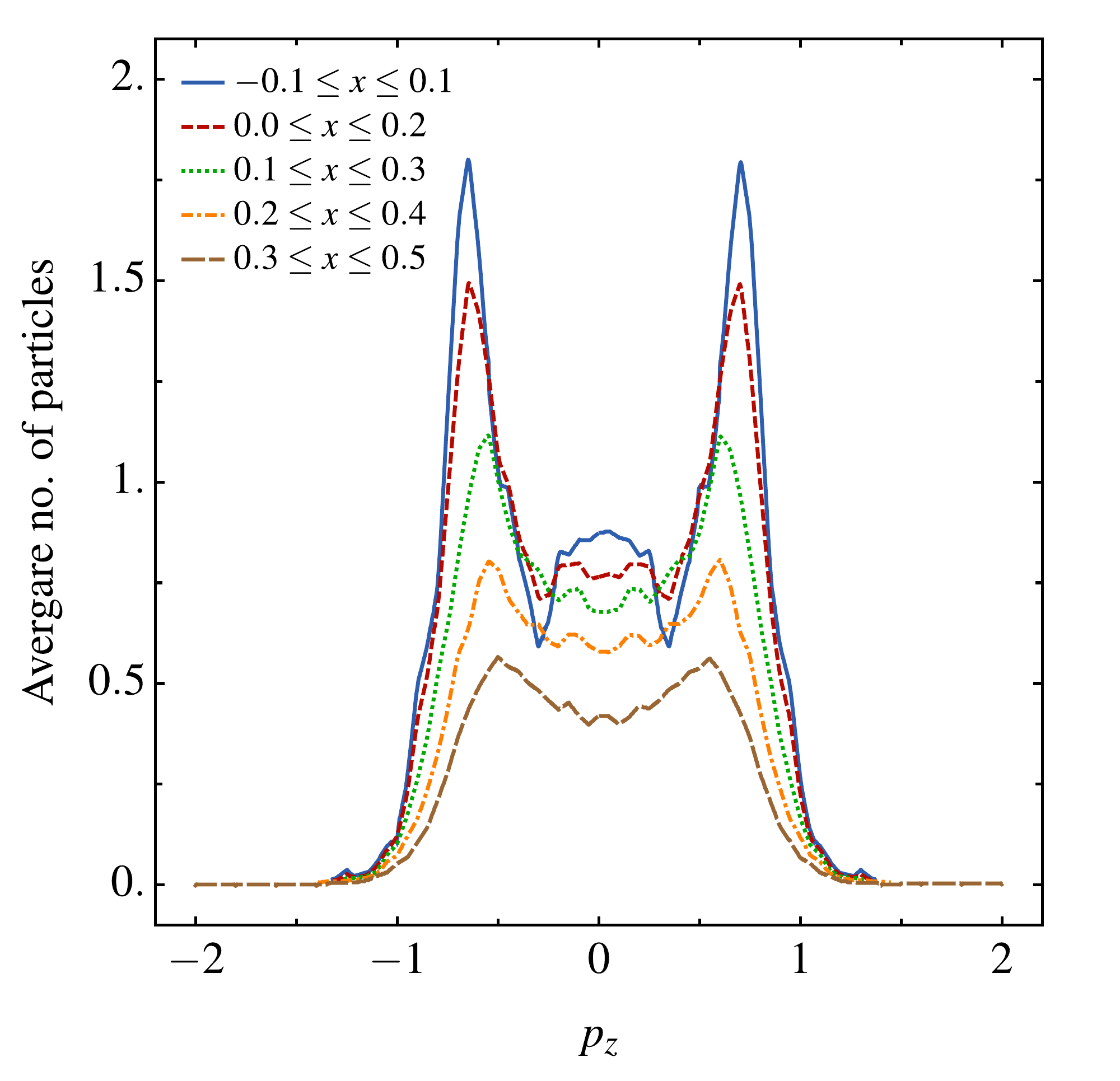}
\caption{The long time Boltzmann distribution $f$ after a kick along $z$ plotted as a function of $p_z$ keeping all other variables $x,y,z,p_x,p_y$ fixed and for different values of $x$. To obtain a nonzero number of atoms in the six dimensional volume we considered a narrow region in phase space  given by the coordinates in the figure and divided it into bins. We plot the number of atoms in each bin averaged over time. It can be seen that $f$ does not resemble a Gaussian thermal function and that the three peaks feature becomes more prominent for $x$ closer to the center. Similar results are found plotting along all other coordinates.} \label{fig:simul2}
\end{figure}

We can ask to what extent the final state of the gas is close to a thermal state. For example, could it be e.g. a product of three different Gaussians (with different temperatures) of the type
\begin{equation}
f \sim e^{-p_x^2/2T_x} e^{-p_y^2/2T_y} e^{-p_z^2/2T_z} \times e^{-V/T} \quad ?
\end{equation}
It is easy to see that this is not possible since it does not satisfy the time independent collisionless Boltzmann equation. In fact we can plot a ``slice'' of $f$ as a function of one of its six coordinates keeping all others fixed as in Fig.~\ref{fig:simul2} which shows a markedly non bell-shaped curve.

In fact, the Gaussian character is only restored upon integration of the other five coordinates of the Boltzmann distribution e.g.:
\begin{equation}
np_x= \int d^3 \mathbf{r} \int dp_y dp_z f(\mathbf{r},\mathbf{p},t \rightarrow \infty)  \propto {\rm e}^{-p_x^2/2mk_BT_x}
\end{equation}
which raises the question of why averages over complex distributions such as those of Fig.~\ref{fig:simul2} lead to a Gaussian profile. We will not consider this question further here, leaving it for further study.

\subsection{Symmetries and sum rule of the distribution}
We can be more quantitative regarding the $\Delta T_i$. We first notice that the quadrupole potential \eqref{eq:potential} is homogeneous of order one, (a potential homogeneous of order $\alpha$ has the property that $V(\lambda {\bf r})=\lambda^\alpha V({\bf r})$). So we can apply the virial theorem which leads to the following relation \cite{Landau1982}:
\begin{equation}
\Delta E = \frac{3}{2} (\Delta T_x + \Delta T_y + \Delta T_z). \label{eq:VT}
\end{equation}

Furthermore, for small kicks we can derive some symmetry considerations and a sum rule.
Defining the matrix $\Theta_{ij}$ as
\begin{equation}
\Delta T_i \equiv \sum_j \Theta_{ij} \frac{q_j^2}{2} \label{eq:Theta}
\end{equation}
where $i,j=x,y,z$, and $q_i$ is the momentum kick along the $ith$ direction, it is possible to show that this matrix is symmetric so that, for small kick momentum, we have $\Theta_{ij}=\Theta_{ji}$.

More generally it is straightforward to show using \eqref{eq:deltaE} and \eqref{eq:VT} that, if the potential is homogeneous of order $\alpha$, the $\Theta_{ij}$ satisfy the constraint
\begin{equation}
 \sum_j \Theta_{ij} = \frac{2 \alpha}{2 + \alpha}.  \label{eq:sum_rule}
 \end{equation}

For potentials with axial symmetry around the $z$ axis, which is the case of the quadrupole trap, the fact that the matrix is symmetric and that in any kick $\Delta T_x=\Delta T_y$ imply that the matrix can be written using only three distinct elements $\theta_{1,2,3}$ as
\begin{equation}
 \Theta=  \begin{pmatrix}
                             \theta_1 & \theta_1 & \theta_2 \\
                             \theta_ 1 & \theta_1 & \theta_2 \\
                             \theta_2 & \theta_2 & \theta_3
                            \end{pmatrix} \label{eq:theta_matrix_axial}  .
\end{equation}
Using the sum rule we find $\theta_2+2\theta_1=\theta_3+2 \theta_2=2/3$ which leaves us with a single unknown parameter. The experimentally measured value $\Delta T_z/q^2_z/2=2/3$ (see point 3. above) implies $\theta_3=2/3$, $\theta_2=0$ and $\theta_1=1/3$, the latter also being in agreement with the measured value.

The quadrupole simulations confirm the experimental observations (1-3) (see Fig.~\ref{fig:simul1}) even though there is a small correction to the experimental values: a slight cooling of the directions transverse to the kick so that $\Theta_{xx}=0.36$ (instead of 1/3) and $\Theta_{xz}=-0.05$. So the observation of point 3., the apparent separability of the $z$ and $x-y$ distributions, seems not to be perfect but rather an excellent approximation. \footnote{We also investigated anisotropic potentials, finding very similar behavior.}

\begin{figure}
\includegraphics[width=0.9\linewidth]{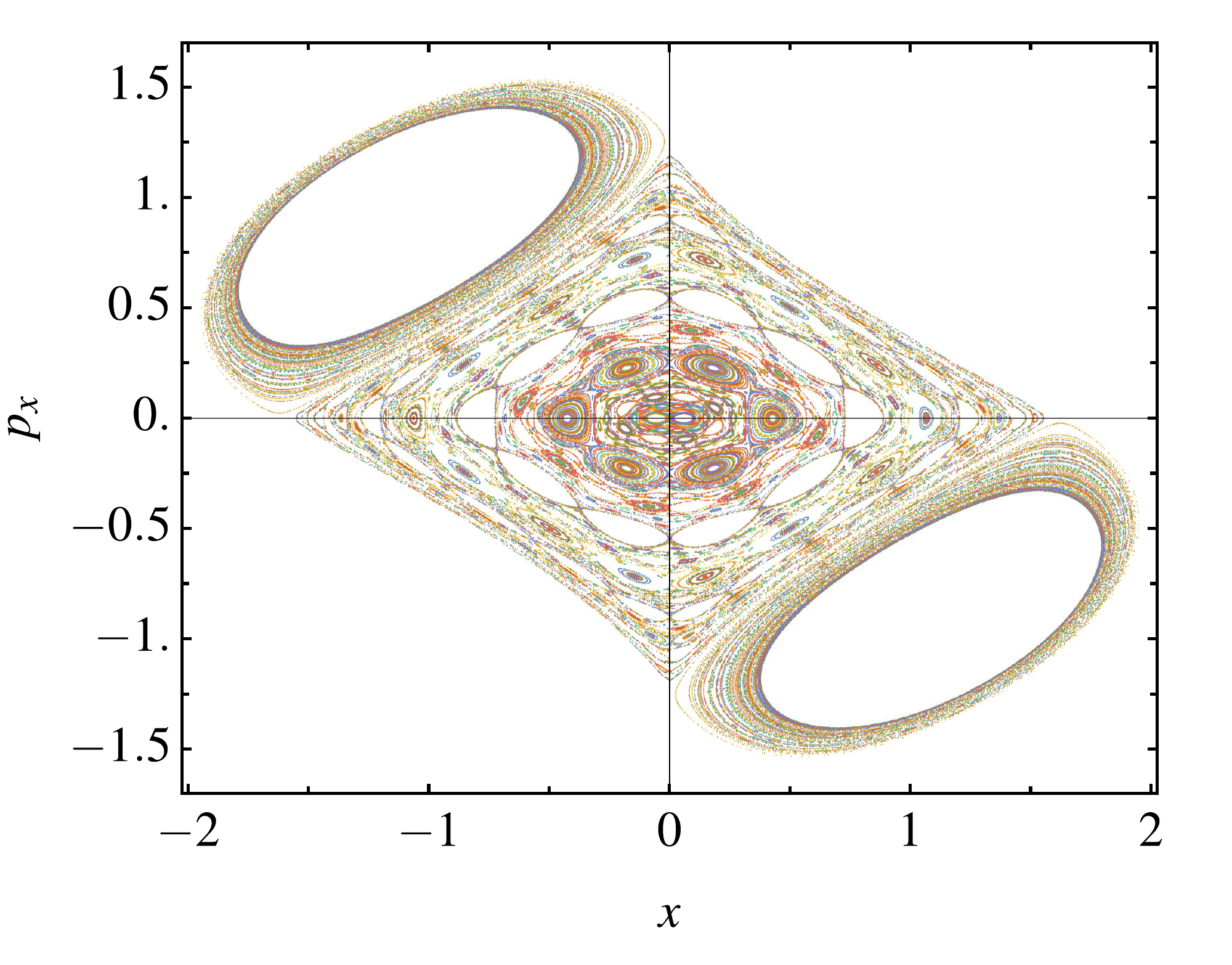}
\caption{\label{Poincare}Poincar\'e map of the quadrupole potential. We study trajectories with the same energy but different initial phase-space coordinates. The values of $x$ and $p_x$ are recorded whenever $z = 0$ and $p_z > 0$. We see the appearance of small islands denoting invariant torii close to which quasi-integrable trajectories evolve, separated by contiguous regions of chaotic dynamics.}
\end{figure}

We can study the gas dynamics by analyzing individual atomic trajectories and then averaging over initial conditions. However the trajectories can be quite difficult to find due to the nonintegrability of the potential. To show this we constructed a Poincar\'e map: in Fig.~\ref{Poincare},  we see that there are both chaotic and quasi-integrable regions. A study of the gas starting from its individual trajectories would be quite complex analytically.
For this reason, it is easier to study not the potential \eqref{eq:potential} but cases which might contain the same physics but in which all or nearly all trajectories are integrable or quasi-integrable. For example let us consider the family of potentials
\begin{equation} \label{eq:general_potential_with_epsilon}
 V_\epsilon(x,y,z) =  \sqrt{x^2 + y^2 +  (1+\epsilon) z^2}.
\end{equation}
When $\epsilon=3$ we get the quadrupole potential (\ref{eq:potential}). But if we take $\epsilon=0$ the potential becomes spherically symmetric and therefore integrable. Alternatively, if $\epsilon \gg 1$ then we are left with a highly confined potential along the $z$ direction (a ``pancake') so that the motion simplifies again and an effective motion in the $x-y$ plane can be studied.


We will begin with the study of the spherical potential in Sec.~\ref{sec:spherical} which, surprisingly, exhibits many of the phenomena of the quadrupole potential, including the anisotropy of the momentum distribution. After this we will analyze the pancake case in Sec.~\ref{sec:pancake}, comparing both of these limits with the quadrupole potential.

\section{Spherical limit} \label{sec:spherical}
The simulations in the quadrupole potential suggest that after perturbing an equilibrium gas along a particular direction, the ensemble average of the momentum widths $\langle p_i^2 \rangle_{t}$ converges to a stationary distribution in the long time limit $t \rightarrow \infty$. In particular, we observed that $\langle p_x^2 \rangle_{\infty} = \langle p_y^2 \rangle_{\infty}$ and in general $\langle p_x^2 \rangle_{\infty} \neq \langle p_z^2 \rangle_{\infty}$.

Calculating the final momentum widths $\langle p_i^2 \rangle_{\infty}$ for a gas of atoms in the quadrupole potential from first principles is difficult without understanding the individual trajectories. Therefore, as mentioned above, it is a natural simplification to consider instead the case where we remove the anisotropy in the quadrupole potential:
\begin{equation} \label{eq:spherical_potential}
 V_{\epsilon=0}(x,y,z) =  \sqrt{x^2 + y^2 +  z^2}=r,
\end{equation}
where $r$ is the radial coordinate (for the rest of this section we will drop the subscript $\epsilon=0$). Na\"ively, one would expect that perturbing a gas along any direction in such an spherical potential will lead to an isotropic distribution at long times: $\langle p_x^2 \rangle_{\infty} = \langle p_y^2 \rangle_{\infty} = \langle p_z^2 \rangle_{\infty}$. However, as we shall see, the final momentum width along the direction of the perturbation will be different to that along perpendicular directions. To anticipate some of the conclusions of this section: this is intuitively plausible: in a spherical potential all three components of angular momentum are conserved, so the motion of each atom is confined to a plane passing through $r=0$ and perpendicular to its angular momentum. The population of each plane is therefore constant during the motion. In thermal equilibrium, this population is the same for all planes but a momentum kick will cause a transfer of atoms between planes, so that the population of each plane will depend on its angle relative to the kick direction. This anisotropy in populations in the distribution is preserved at long times again due to conservation of angular momentum and translates into different final temperatures along the different directions.

\subsection{Averages over the motion in planes} \label{sec:Planes}
\begin{figure}
\includegraphics[width=0.9\linewidth]{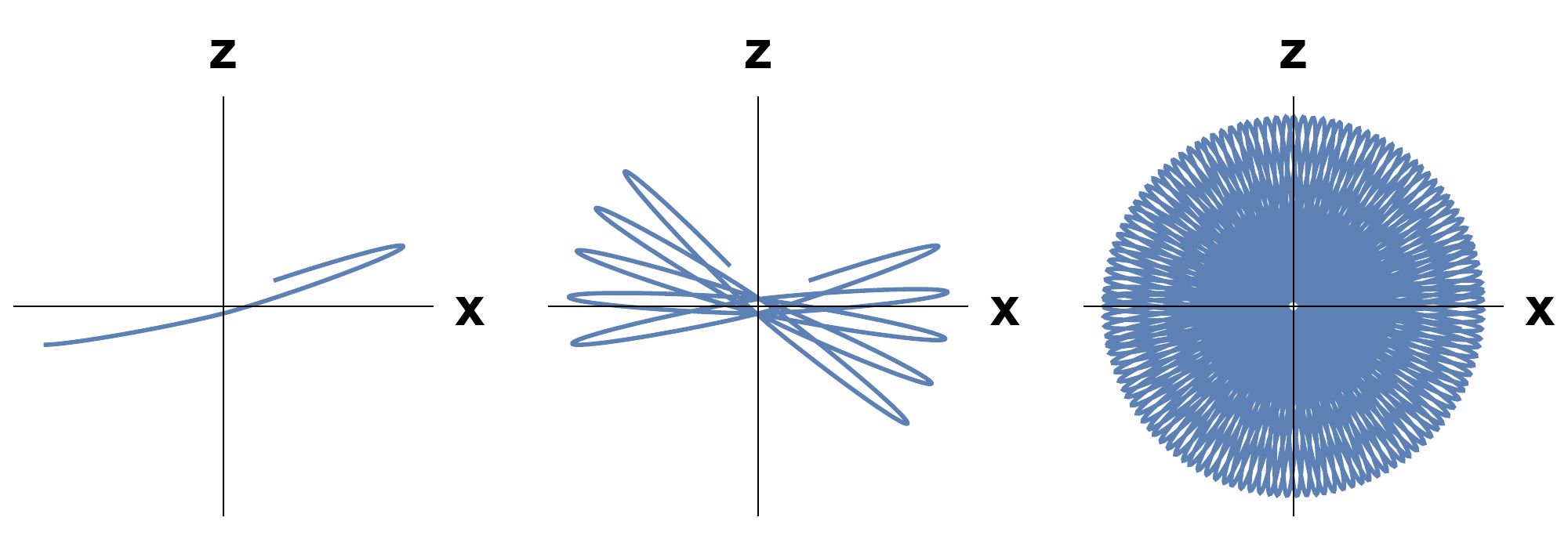}
\caption{\label{Bertrand}Orbit of atom in a plane with a central potential $V(r)=r$ after increasingly long times from left to right. Since the trajectory never closes, according to Bertrand's theorem, it fills the annular region between $r_{\rm max}$ and $r_{\rm min}$ in an isotropic, dense fashion as $t \rightarrow \infty$.}
\end{figure}

With a particle in a central field \cite{Landau1982,Goldstein2002}, the trajectory stays on the plane perpendicular to its angular momentum $\mathbf{L}$ which includes the origin $r=0$. Using polar coordinates $(r,\theta)$ for the plane, the energy $E$ is given by the usual expression:
\begin{equation}
 E = \frac{1}{2} \left( \dot{r}^2 + r^2 \dot{\theta}^2 \right) + V(r) = \frac{1}{2} \left(  \dot{r}^2 + \frac{L^2}{ r^2} \right) + V(r)
\end{equation}
where $L =|\mathbf{L}|=  r^2 \dot{\theta} = \text{constant}$. 
In a potential such as (\ref{eq:spherical_potential}), the motion is confined between two values of the radial coordinate $r_{\rm min} \le r \le r_{\rm max}$ which are solutions of $\dot{r} = 0$. 
During the time in which $r$ varies from $r_{\rm max}$ to $r_{\rm min}$ and back, the radius vector turns through an angle $\Delta \theta$. 
The condition for the path to be closed is that this angle should be a rational fraction of $2 \pi$, i.e. that $\Delta \theta = 2 \pi m / n$, where $m$ and $n$ are integers. But according to Bertrand's theorem \cite{Goldstein2002} the only central potentials for which all paths are closed are Kepler's ($\propto -\frac{1}{r}$) and the harmonic potential ($ \propto r^2$). For all other potentials (and excluding the particular case of trajectories with zero angular momentum), the trajectory will behave as in Fig.~\ref{Bertrand}: it will become dense everywhere, filling the allowed annulus region isotropically so that the orbital density is only a function of the radius $r$ as the propagation time tends to infinity.

Using Bertrand's theorem, we would like to analyze the long time behavior of trajectories, in particular the time averages of different quantities. For a quantity $A(t)$, the time average of a quantity $\overline{A}$ is defined as, cf.~(\ref{eq:phase_space_average}):
\begin{equation}
 \overline{A} \equiv \lim_{t \rightarrow \infty} \frac{1}{t} \int_0^{t} A(t^\prime) d t^\prime. \label{eq:time_average}
\end{equation}
We can convert the time average to one over the orbital density discussed above by a change of variables. We immediately conclude that, since Bertrand's theorem implies that the orbital density is isotropic, so will the time average also be:
\begin{eqnarray}
 \overline{x^2} & = & \overline{y^2} \label{eq:x2_eq_y2}
\\
 \overline{p_x^2} & = & \overline{p_y^2}. \label{eq:px2_eq_py2}
\end{eqnarray}
We will use this fact to calculate $\langle p_i^2 \rangle_{\infty}$ for a gas of atoms.

\subsection{Calculation of momentum averages in terms of integrals of planes}
\label{sec:IsotropicDerivation}
Although our purpose is to study the potential $V(r)=r$ as a limiting case of the family \eqref{eq:general_potential_with_epsilon}, it is straightforward to consider in this section a more general potential than (\ref{eq:spherical_potential}), namely
\begin{equation} \label{eq:general_isotropic_potential}
 V(r) = r^\alpha
\end{equation}
with $0<\alpha \neq 2$. This will allow us to examine qualitatively different behavior as a function of $\alpha$. The case $\alpha = 2$ corresponds to the isotropic harmonic potential for which in general (\ref{eq:x2_eq_y2}) and (\ref{eq:px2_eq_py2}) are not true. For $\alpha=1$ we recover (\ref{eq:spherical_potential}).

For a gas in an spherical potential, the atoms belonging to the same plane in coordinate space are also confined to the same plane in momentum space making each plane an independent system. So our strategy will be to treat the motion in each plane separately and then add over all of them at the end. For this we choose a coordinate system (see Appendix~\ref{sec:coordinates}) where two of the coordinates (the angles $\theta$ and $\phi$) define the plane, and the remaining four correspond to the in-plane coordinates ($u$ and $v$) and momenta ($p_u$ and $p_v$). Then we can write the total energy as
\begin{equation}
 \langle E \rangle = \int_0^{\pi} d \phi \int_0^{\pi} d \theta \langle E \rangle _{\rm plane}.
 \label{eq:energy_energy_plane_relationship}
\end{equation}
where $\langle E \rangle _{\rm plane}$ is the average energy of all the planes lying between $\theta$ and $\theta+d\theta$, $\phi$ and $\phi+d\phi$.
Even though the probability density $f(\mathbf{r},\mathbf{p},t)$ is a function of time, the energy of each atom is constant in time as the potential is time-independent and there is no exchange of energy between the atoms, so the average energy is also a constant. Therefore if we know the probability density $f(\mathbf{r},\mathbf{p},t)$ at any one time, we will know the average energy for all time. This allows us to calculate the final momentum widths $\langle p_i^2\rangle_\infty$ from the distribution of energies at $t=0$ after the initial momentum kick.

Since the class of potentials (\ref{eq:general_isotropic_potential}) is homogeneous of order $\alpha$ we use the virial theorem,
\begin{equation}
 \overline{K} =  \frac{\alpha}{2} \overline{V},
\end{equation}
where $K$ is the kinetic energy and the averages are over time as in \eqref{eq:time_average}. Note that the virial theorem is valid both for each atom individually as well as for the entire gas. If we assume that at long times, when the gas has reached a steady state, the ergodic hypothesis applies for such systems, we can replace the time average with the ensemble average
\begin{equation} \label{eq:virial_theorem_power_potentials}
 \langle K \rangle = \frac{\alpha}{2} \langle V \rangle.
\end{equation}

As each plane is a closed individual system, (\ref{eq:virial_theorem_power_potentials}) also applies to
\begin{equation} \label{eq:plane_virial_theorem_power_potentials}
 \langle K \rangle_{\rm plane} = \frac{\alpha}{2} \langle V \rangle_{\rm plane},
\end{equation}
and using (\ref{eq:plane_virial_theorem_power_potentials}), $\langle E \rangle _{\rm plane}$ can be written as
\begin{eqnarray}
 \langle E \rangle _{\rm plane} & = & \langle K \rangle _{\rm plane} + \langle V \rangle _{\rm plane} \nonumber
\\
 & = & \frac{2+\alpha}{\alpha} \langle K \rangle _{\rm plane} \nonumber
\\
 & = & \frac{2+\alpha}{2 \alpha} \left( \langle p_u^2 \rangle + \langle p_v^2 \rangle \right).
\end{eqnarray}
According to Bertrand's Theorem, Kepler's potential $V(r) = -\frac{k}{r}$ and radial harmonic oscillator $V(r) = \frac{1}{2} k r^2$ are the only two types of central force potentials where all bound orbits are also closed orbits. Therefore, if we restrict ourselves to cases where $0< \alpha \ne 2$ where almost all orbits are open (except for the circular orbit), we see that $\langle p_u^2 \rangle = \langle p_v^2 \rangle$ as $t \rightarrow \infty$ so that, following the argument of Sec.~\ref{sec:Planes},
\begin{equation} \label{eq:energy_plane_pu2_relation}
 \langle p_u^2 \rangle = \langle p_v^2 \rangle = \frac{ \alpha}{2 + \alpha} \langle E \rangle _{\rm plane}.
\end{equation}
We can now express the the averages of $p_x^2$, $p_y^2$ and $p_z^2$ through $\langle E \rangle _{\rm plane}$ as shown in Appendix~\ref{sec:momenta} (assuming that the final distribution does not depend on $\phi$)
\begin{eqnarray}
T_{x,y}= \langle p_{x,y}^2 \rangle & = & \frac{ \alpha \pi}{2(2+\alpha)} \int_0^{\pi} d \theta \langle E \rangle _{\rm plane} (1 + \sin^2 \theta) \label{eq:average_pxy2_final}
\\
T_z= \langle p_z^2 \rangle & = & \frac{\alpha \pi}{2+\alpha} \int_0^{\pi} d \theta \langle E \rangle _{\rm plane} \cos^2 \theta. \label{eq:average_pz2_final}
\end{eqnarray}
It remains now to calculate $\langle E\rangle_{\rm plane}$ as a function of $\theta$ and $\phi$ after the momentum kick.
\subsection{Momentum Kick}
\begin{figure}[!t]
\centering
  \includegraphics[width=0.9\linewidth]{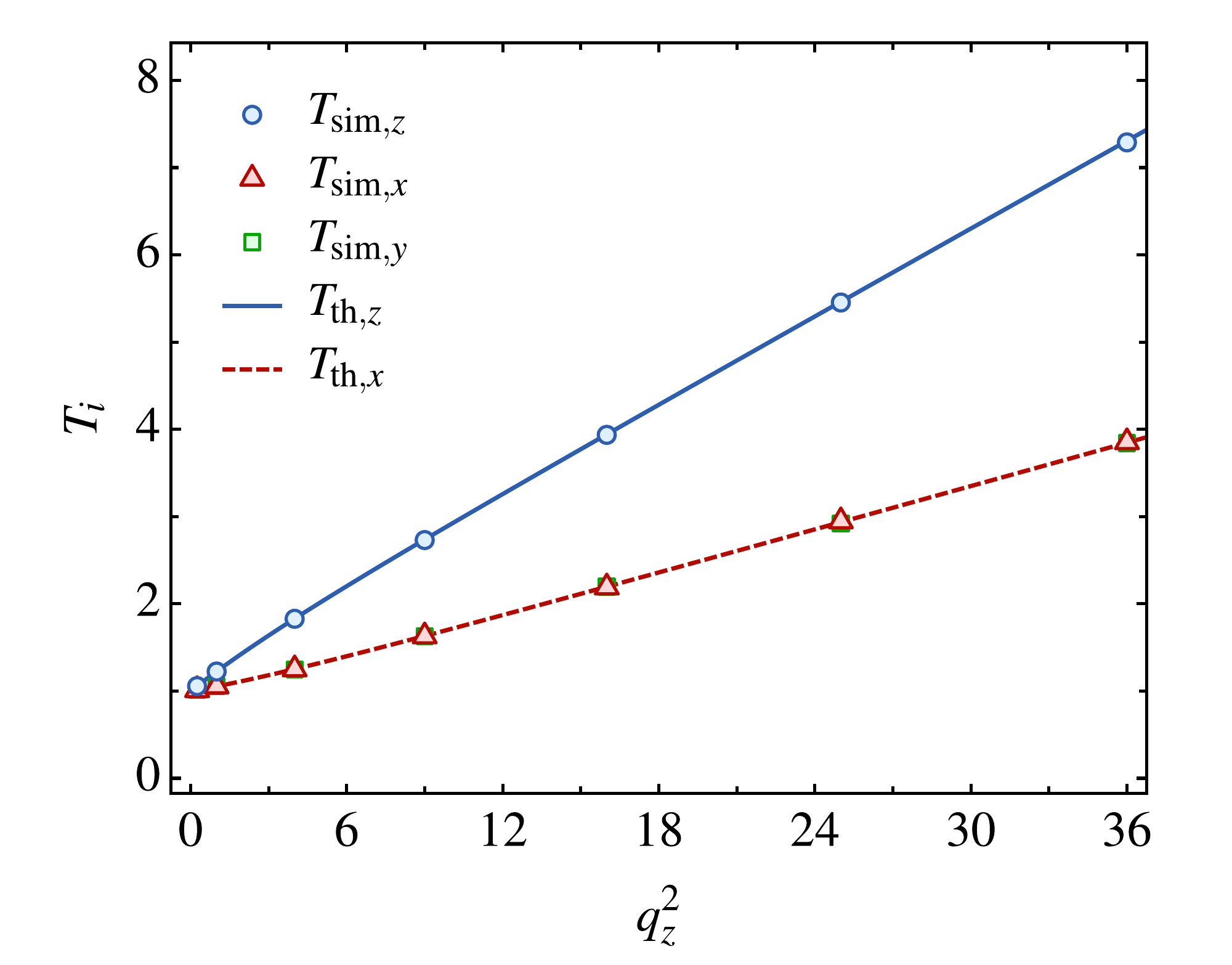}
\caption{Comparison of the simulation results for $T_i=\langle p_i^2 \rangle_{t\rightarrow\infty}$ ($i=x,y,z$) with the analytical predictions (\ref{eq:average_pxy2_analytical}) and (\ref{eq:average_pz2_analytical}) for an isotropic potential (\ref{eq:general_isotropic_potential}) with $\alpha=1$ for different kick strengths $q_z$ along the $z$-direction. Note that the predicted $\langle p_x^2 \rangle_\infty$ and $\langle p_y^2 \rangle_\infty$ are identical.}
\label{Figure:spherical_theory_vs_simulation}
\end{figure}

We perturb the Maxwell-Boltzmann distribution in a potential given by (\ref{eq:general_isotropic_potential}) at $t=0$ by applying a momentum kick $q_z$ along the $z$-direction. The resulting initial distribution at temperature $k_BT_0=1$ is:
\begin{eqnarray}
\label{eq:momentum_kick_spherical_distribution}
\lefteqn{f(\mathbf{r},\mathbf{p},t=0) =}\nonumber \\ && A \exp \left( -  \frac{p_x^2 + p_y^2 + (p_z - q_z)^2}{2} -  r^\alpha  \right)
\end{eqnarray}
where
\begin{equation}
A = \frac{3}{8 \sqrt{2} \pi^{5/2}\Gamma \left( \frac{3+\alpha}{\alpha} \right)}.
\end{equation}
If we transform (\ref{eq:momentum_kick_spherical_distribution}) using (\ref{eq:polar_plane_transformation}), we get:
\begin{eqnarray} \label{eq:momentum_kick_spherical_distribution_polar_plane}
\lefteqn{f = A \exp \left( - \frac{q_z^2}{2 } \right) } \nonumber\\ &&\exp \left( -  r^{\alpha} \right) \exp \left( - \frac{p_r^2 - 2 p_r q_z \cos \theta \sin \alpha_p}{2} \right).
\end{eqnarray}
Using (\ref{eq:momentum_kick_spherical_distribution_polar_plane}), we can calculate $\langle V \rangle _{\rm plane}=\langle r^\alpha\rangle$, $\langle K \rangle _{\rm plane}=\langle p_r^2\rangle/2m$, and finally $\langle E \rangle _{\rm plane}= \langle V \rangle _{\rm plane} + \langle K \rangle _{\rm plane}$ as follows:
\begin{eqnarray}
\!\!\!\!\!\langle V \rangle _{\rm plane} (t=0) & &= \frac{3 |\cos \theta|}{ \alpha \pi} {\rm e}^{ - \frac{q_z^2}{2}} {\rm e}^{ \frac{q_z^2 \cos^2 \theta}{4} } \nonumber
\\
 & & \times \left[  I_1 \left( \frac{q_z^2 \cos^2 \theta}{4} \right) \frac{q_z^2 \cos^2 \theta}{4}\right.\nonumber
 \\
 & &\left.+ I_0 \left( \frac{q_z^2 \cos^2 \theta}{4} \right) \left(\frac{1}{2} + \frac{q_z^2 \cos^2 \theta}{4}\right) \right], \label{eq:plane_pe_momentum_kick}
\end{eqnarray}
\begin{eqnarray}
&& \langle K \rangle _{\rm plane} (t=0)  =  \frac{|\cos \theta|}{8 \pi} \exp \left( - \frac{q_z^2}{2} \right) \exp \left( \frac{q_z^2 \cos^2 \theta}{4} \right) \nonumber
\\
 & & \times \left[ q_z^2 I_1 \left( \frac{q_z^2 \cos^2 \theta}{4} \right) \cos^2 \theta (4+ q_z^2 \cos^2 \theta)\right. \nonumber
\\ & & \left.+ I_0 \left( \frac{q_z^2 \cos^2 \theta}{4} \right) (6+ 6 q_z^2 \cos^2 \theta + q_z^4 \cos^4 \theta) \right],
\label{eq:plane_ke_momentum_kick}
\end{eqnarray}
where $I_0$ and $I_1$ are modified Bessel functions of the first kind.
Since $\langle E \rangle _{\rm plane}$ does not change with time we can use this to obtain the $\langle p_i^2\rangle$ at $t\rightarrow\infty$ via Eqs.~(\ref{eq:average_pxy2_final},\ref{eq:average_pz2_final}). For $\alpha=1$ the resulting expressions read
\begin{eqnarray}
\langle p_x^2 \rangle &=& \frac{q_z^2}{12} + \frac{5}{6} + \frac{1}{2q_z^2} - \frac{\sqrt{2}}{2q_z^3} F \left( \frac{q_z}{\sqrt{2}} \right),\label{eq:average_pxy2_analytical}\\
\langle p_z^2 \rangle &=& \frac{q_z^2}{6} + \frac{4}{3} - \frac{1}{q_z^2} + \frac{ \sqrt{2}}{q_z^3} F \left( \frac{q_z}{\sqrt{2}} \right),\label{eq:average_pz2_analytical}
\end{eqnarray}
where $F$ is the Dawson function. In Fig.~\ref{Figure:spherical_theory_vs_simulation} we show the excellent agreement of the simulations with these analytical predictions.

For a small momentum kick, we can find some illuminating expressions. Expanding $\langle E \rangle _{\rm plane}$ about $q_z = 0$ up to $\mathcal{O}(q_z^2)$ we obtain from (\ref{eq:average_pxy2_final}) and (\ref{eq:average_pz2_final})
%
\begin{eqnarray}
T_{x,y} & \approx & 1 + \frac{5 \alpha - 2}{20 (2 + \alpha)} q_z^2 \label{eq:px2_approx}
\\
T_z & \approx & 1 + \frac{2 + 5 \alpha}{10 (2 + \alpha)} q_z^2. \label{eq:pz2_approx}
\end{eqnarray}
For the case $\alpha=1$ we find
\begin{eqnarray}
T_{x,y} & \approx & 1 + \frac{1}{20} q_z^2 \Rightarrow \Delta T_{x,y}=\frac{1}{10} \Delta E, \label{eq:sphericaltemperatures1}
\\
T_z & \approx & 1 + \frac{7}{30} q_z^2 \Rightarrow \Delta T_{x,y}=\frac{7}{15} \Delta E. \label{eq:sphericaltemperatures2}
\end{eqnarray}
which satisfies the virial theorem \eqref{eq:VT}. Comparing with the quadrupole experiment (point 3. above) where $\Delta T_{x,y}=0$ and $\Delta T_z = 2/3 \Delta E$, we see that the spherical case leads to some increased heating in the $xy$ plane although small.

In terms of the matrix $\Theta_{ij}$ from \eqref{eq:Theta}, for a spherically symmetric case we can show that
\begin{equation}
 \Theta_{ij} =  \begin{pmatrix}
                             \theta_1 & \theta_2 & \theta_2 \\
                             \theta_2& \theta_1 & \theta_2 \\
                             \theta_2 & \theta_2 & \theta_1
                            \end{pmatrix} \label{eq:theta_matrix_spherical}
\end{equation}
so that e.g. $\Delta T_x=\theta_1 q^2_x/2$ and $\Delta T_x=\theta_2 q^2_y/2$. As before, using the sum rule, we find that $\theta_1+2\theta_2=2/3$ so that the matrix depends only on a single unknown parameter.
Then (\ref{eq:px2_approx}) and (\ref{eq:pz2_approx}) imply that
\begin{equation}
\theta_1=\frac{2 + 5 \alpha}{5 (2 + \alpha)} \mbox{ and } \theta_2=\frac{5 \alpha - 2}{10 (2 + \alpha)}
\end{equation}
which satisfy the sum rule \eqref{eq:sum_rule}. For the case $\alpha=1$ (\ref{eq:spherical_potential}) we get $\theta_1=7/15$ and $\theta_2=1/10$.


\subsection{Heating and cooling of transverse directions}
These results allow us to answer an interesting question: if we kick the gas along a direction, do the transverse directions heat or cool?

For an interacting gas, we know collisions will distribute the energy along all directions, hence the transverse directions will be heated by the same amount as the kicked direction. For an ideal gas in e.g.\ a harmonic potential, the transverse directions will not be affected.

Using (\ref{eq:px2_approx}) and (\ref{eq:pz2_approx}) we see that, for a noninteracting gas in a spherical potential of the form (\ref{eq:general_isotropic_potential}), we can have different types of behaviour (up to $\mathcal{O}(q_z^2)$) for the transverse directions:

\begin{itemize}
 \item for $\alpha < \frac{2}{5}$: cooling;
 \item for $\alpha = \frac{2}{5}$: no change;
 \item for $\alpha > \frac{2}{5}$: heating.
\end{itemize}

This surprising result tells us that it is possible in some cases to {\it cool} the gas along some directions while heating it up along others. In fact, as we will see later the quadrupole potential is of this type: it cools along the $x$ and $y$ directions if kicked along $z$. Nevertheless, the spherical potential, which most closely resembles it, with $\alpha=1$, behaves more conventionally since it heats up.

\subsection{Population redistribution due to kick}
We would like to gain some insight into why the final momentum widths are different $\langle p_x^2 \rangle = \langle p_y^2 \rangle \ne \langle p_z^2 \rangle$ for $q_z \ne 0$.

We can rewrite \eqref{eq:average_pxy2_final} using the fact that the total energy of the gas $E_{\rm total}(q=0)+\Delta E$ with $\Delta E$ given by \eqref{eq:deltaE}, can be expressed as the sum of the plane energies:
\begin{equation}
E_{\rm total}(q=0)+\frac{q_z^2}{2}=\int_0^\pi d \phi \int_0^{\pi} d \theta \langle E \rangle _{\rm plane}= \pi \int_0^{\pi} d \theta \langle E \rangle _{\rm plane}
\end{equation}

The term $E_{\rm total}(q=0)$ can be easily found from the $q_z=0$ limit of \eqref{eq:plane_pe_momentum_kick} and \eqref{eq:plane_ke_momentum_kick}. It follows that:
\begin{eqnarray}
\langle p_{x,y}^2 \rangle  &= & \frac{ \alpha \pi}{2(2+\alpha)} \int_0^{\pi} d \theta \langle E \rangle _{\rm plane} (1 + \sin^2 \theta) \nonumber \\
&=&\frac{ \alpha}{2(2+\alpha)} \left( \frac{q^2}{2 } +\frac{6 + 3 \alpha}{2 \alpha } +\pi  \int_0^{\pi} d \theta \langle E \rangle _{\rm plane} \sin^2 \theta \right) \nonumber \\
& \stackrel{\alpha=1}{=}& \frac{1}{6} \left( \frac{q_z^2}{2 } + \frac{9}{2} \right) + \frac{ \pi}{6} \int_0^{\pi} d \theta \sin^2 \theta \langle E \rangle _{\rm plane}. \label{eq:average_px2_separate_a_1}
\end{eqnarray}

\begin{figure}[!htb]
\centering

\includegraphics[width=0.9\linewidth]{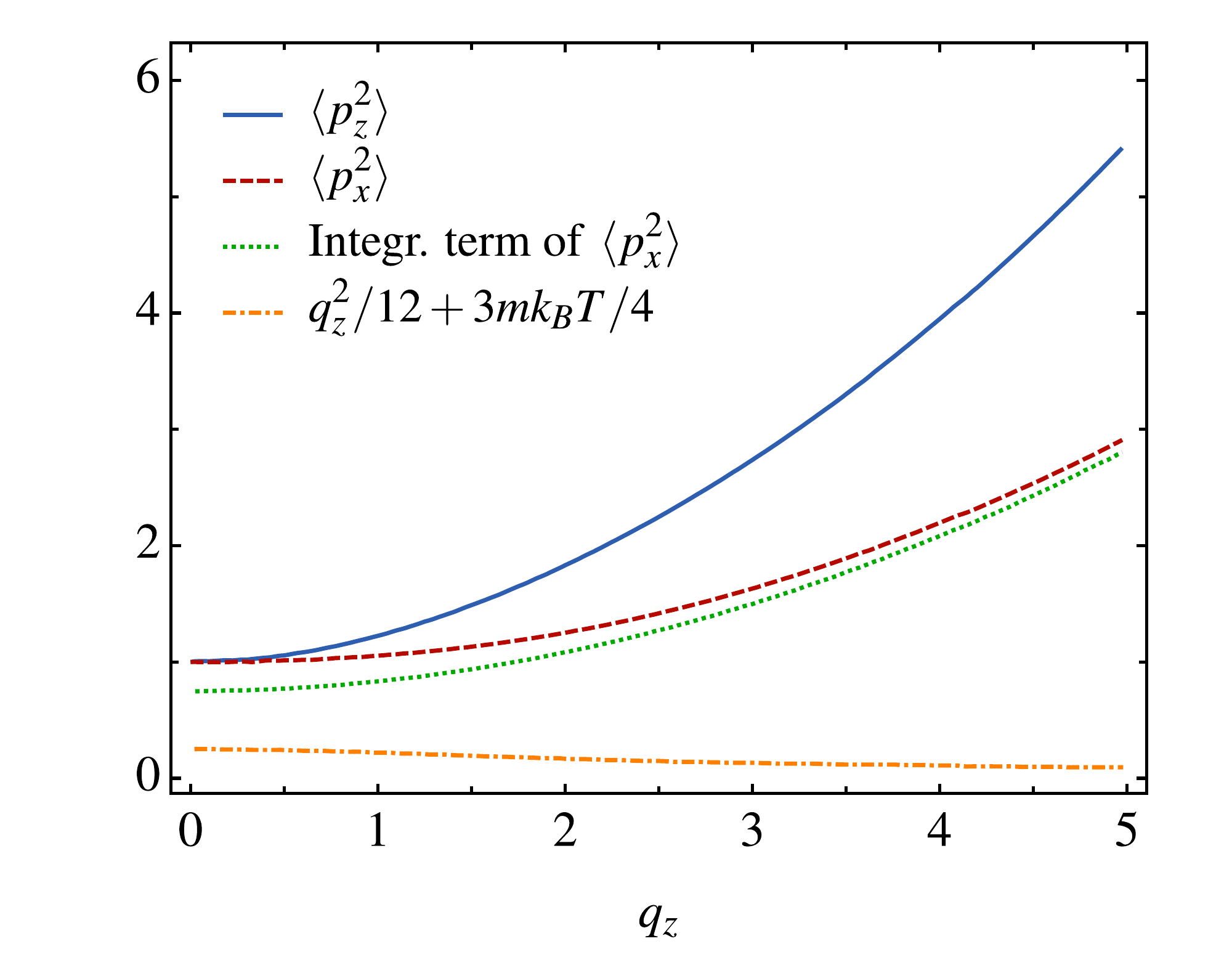}
\caption{Comparing the different terms of $\langle p_x^2 \rangle$ in \eqref{eq:average_px2_separate_a_1} with $\langle p_z^2 \rangle$ \eqref{eq:average_pz2_analytical} for different values of momentum kick $q_z$ with $m=k_B T=1$.}
\label{Figure:momentum_width_terms_vs_qz}
\vspace{0.4cm}
\includegraphics[width=0.9\linewidth]{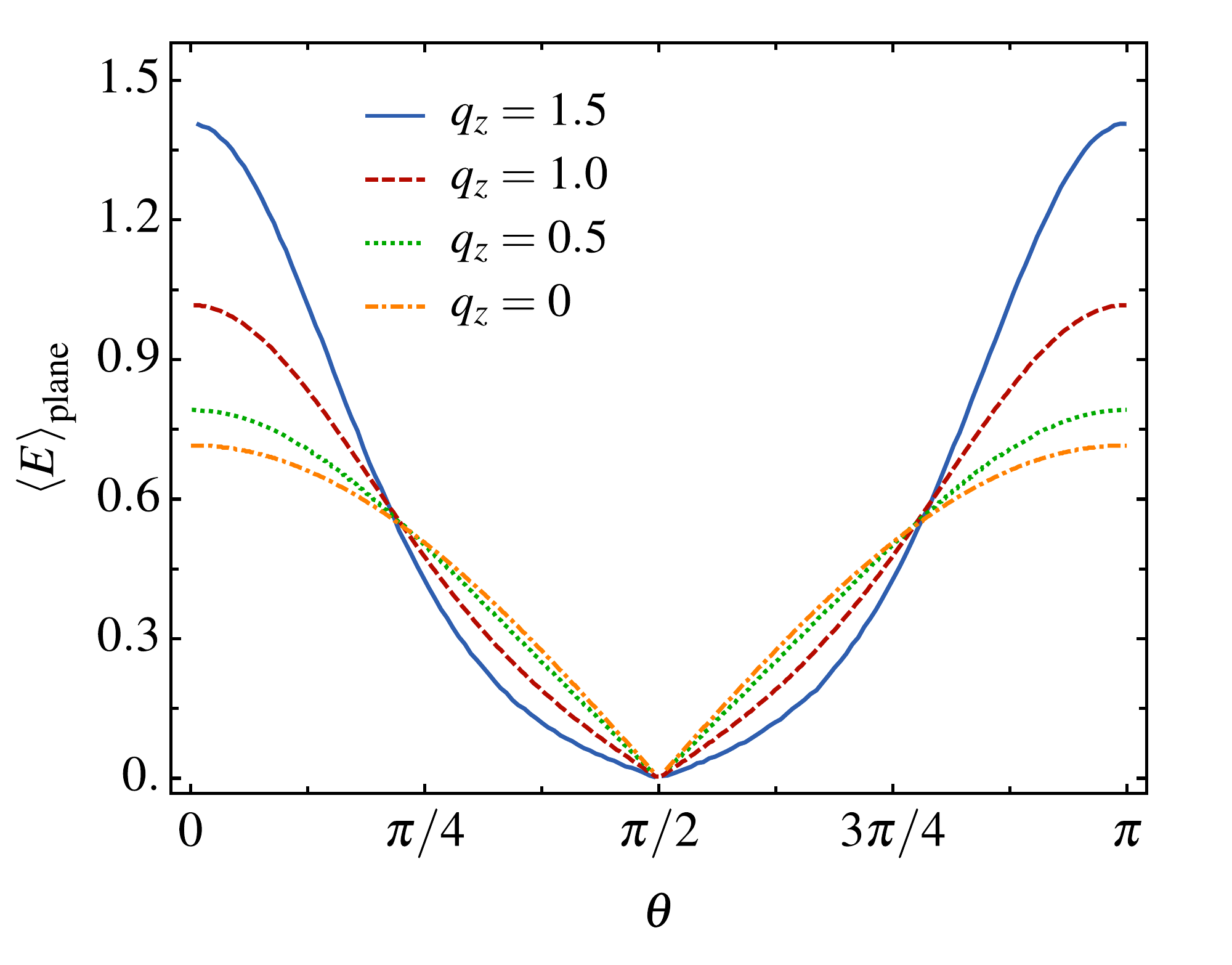}
\caption{Using Eqs.~\eqref{eq:plane_pe_momentum_kick} and \eqref{eq:plane_ke_momentum_kick} to  plot $\langle E \rangle _{\rm plane} (\theta)$ for different values of momentum kick $q_z$.}
\label{Figure:energy_plane_vs_theta_momentum_kick}
\vspace{0.4cm}
\includegraphics[width=0.9\linewidth]{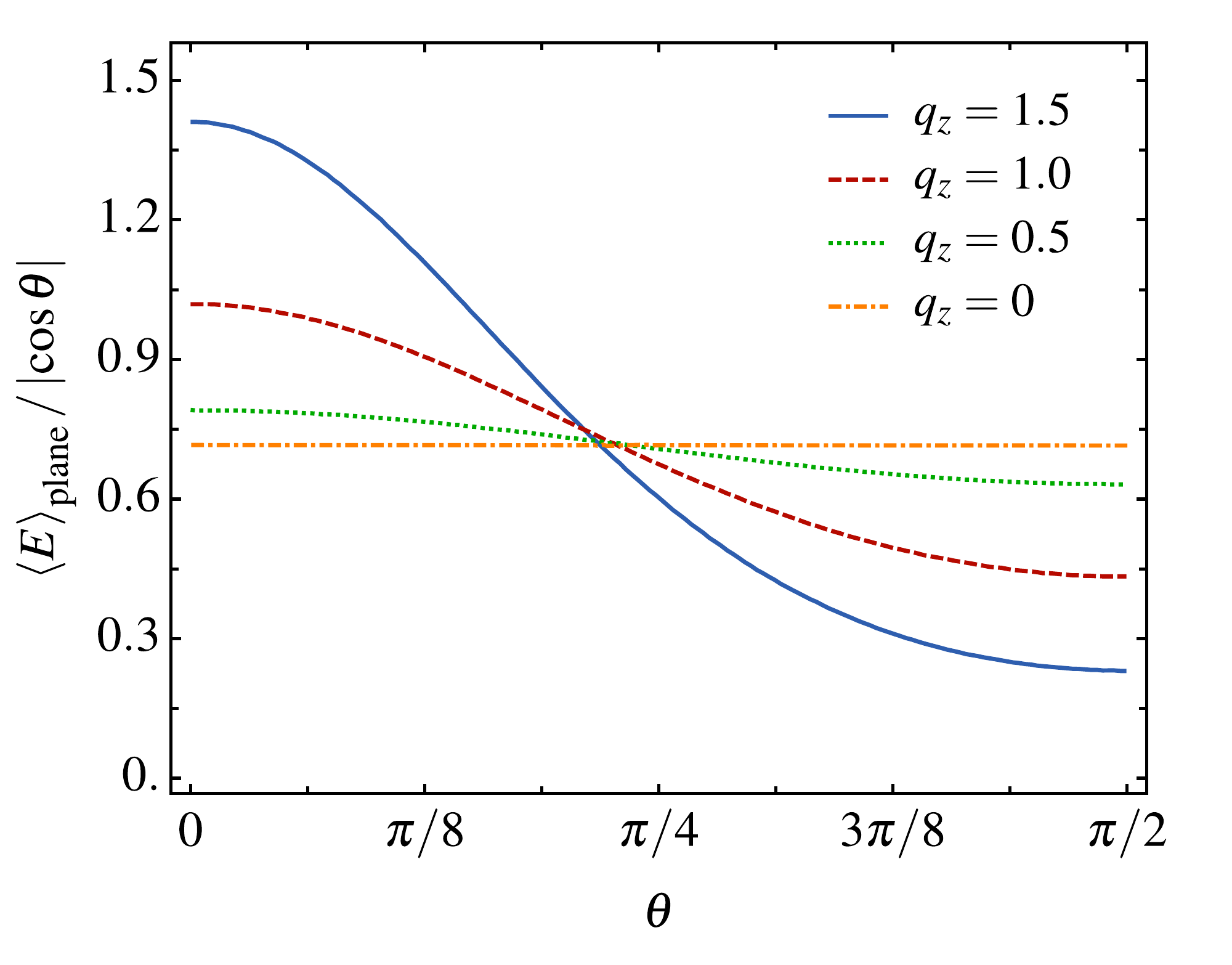}
\caption{Using Eqs.~\eqref{eq:plane_pe_momentum_kick} and \eqref{eq:plane_ke_momentum_kick} to  plot $\langle E \rangle _{\rm plane} (\theta)/|\cos \theta|$ for different values of momentum kick $q_z$.}
\label{Figure:energy_plane_over_cos_vs_theta_momentum_kick}

\end{figure}

We can study how each of the terms in $\langle p_x^2 \rangle$ varies with $q_z$. In Fig.~\ref{Figure:momentum_width_terms_vs_qz}, we can see that the contribution of the integral term of \eqref{eq:average_px2_separate_a_1} is small compared to the $q_z^2$ term and becomes less important as $q_z$ increases.


To understand why the integral term becomes small, we can investigate how $\langle E \rangle _{\rm plane}$ changes as a function of $\theta$ for different values of $q_z$. From Fig.~\ref{Figure:energy_plane_vs_theta_momentum_kick}, we can see that the value of $\langle E \rangle _{\rm plane}$ near $\theta=0$ and $\theta=\pi$ increases with increasing $q_z$ and the opposite happens near $\theta = \pi / 2$. As the integrand multiplies this factor by $\sin^2 \theta$ which is $0$ at $\theta = 0, \pi$ and peaks at $\theta = \pi/2$ the integral will decrease as $q_z$ increases.

To make it even clearer, it is useful to plot not $\langle E \rangle_{\rm plane}$ but $\langle E \rangle_{\rm plane}/|\cos \theta|$ which removes the effect of the Jacobian \eqref{eq:jacobian_cartesian_plane} which simply accounts for the variation of the density of planes as a function of $\theta$, leaving us with the change in plane energy as a result of the kick.

From Fig.~\ref{Figure:energy_plane_over_cos_vs_theta_momentum_kick}, we can see that when there is no momentum kick, the energy of all the planes are the same. When we apply a momentum kick along the $z$-axis, planes lying along that direction ($\theta = 0$ or $\pi$) gain energy whereas directions close $\theta = \pi /2$ lose it. This means that, when we project the energy of each plane to obtain the momentum widths, $\langle p_z^2 \rangle > \langle p_x^2 \rangle$.

\begin{figure}[!t]
  \centering
  \includegraphics[width=0.9\linewidth]{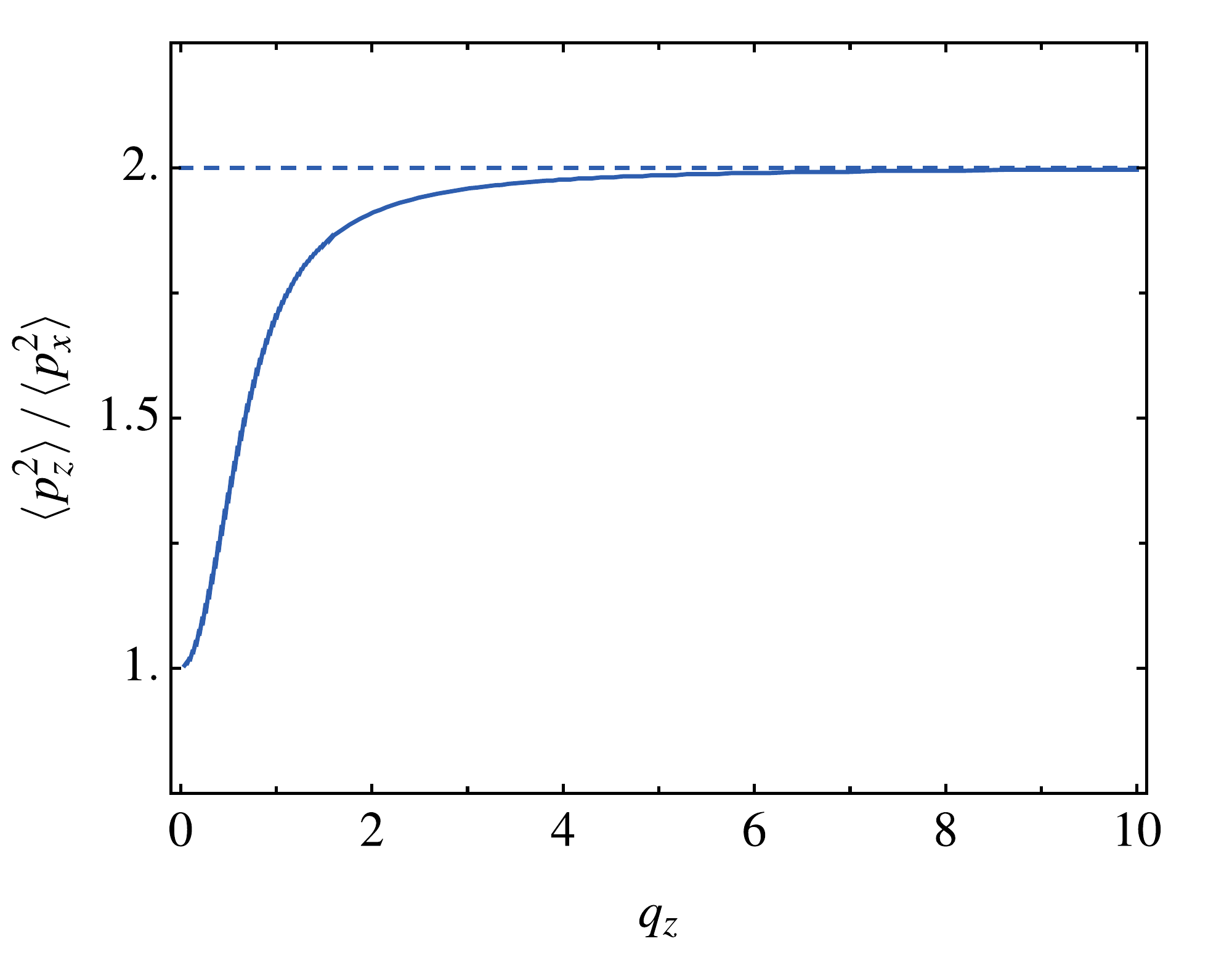}
  \caption{Using Eqs.~\eqref{eq:plane_pe_momentum_kick} and \eqref{eq:plane_ke_momentum_kick} to  plot the ratio between $\langle p_z^2 \rangle$ and $\langle p_x^2 \rangle$ for different values of momentum kick $q_z$.}
  \label{Figure:momentum_widths_ratio}
\end{figure}

We can also see that as $q_z \rightarrow \infty$, $\langle E \rangle _{\rm plane} / |\cos \theta|$ is only non-zero at $\theta = 0$ and $\pi$ which explains the momentum widths ratio constraint derived in \eqref{eq:pz2_px2_ratio_constraint} (note that Fig.~\ref{Figure:momentum_widths_ratio} agrees with the ratio constraint).

\subsection{Memory loss in isotropic potentials}
A natural question arising from the study of this section is whether a gas can remember in which direction it was kicked after a long time has passed. For example, we could start with a gas in thermal equilibrium in an isotropic potential, i.e.\ a spherically (3D) or circularly (2D) symmetric potential, apply a momentum kick along an arbitrary direction and wait for a very long time. Is the final gas distribution anisotropic? I.e.\ does it preserve a memory of the direction of the kick?

In a collisional gas, the extra energy from the momentum kick is redistributed along all directions equally, leading to isotropic heating and therefore a loss of memory.

A non-interacting gas in a harmonic oscillator preserves this memory because its center of mass oscillates along the kick direction indefinitely.

However, quite surprisingly, a non-interacting gas in a non-separable potential can also preserve it due to the existence of integrals of motion which encode the direction. For example in a 3D spherical  potential the memory is associated with the three components of angular momentum $L_{x,y,z}$ being integrals of the motion, as we have seen.

An interesting question is: can there be memory loss with no interactions and a non-separable potential? Unexpectedly the answer is yes: for example a gas in a 2D circular symmetric potential has $\langle p_x^2 \rangle = \langle p_y^2 \rangle$  due to Bertrand's theorem, so memory is lost (excluding harmonic and Kepler's potential). There is only a single component of angular momentum so the direction cannot be encoded in the integrals of the motion.
After the kick the extra energy is redistributed to all directions, the ``orbit density'' becomes isotropic as $t \rightarrow \infty$ which leads to loss of memory. This macroscopic loss of information is due to ergodicity of the individual trajectories rather than to collisions. Of course, microscopically the memory is preserved since, if we reversed the momenta of all atoms at the same time, we could recover the initial kicked distribution.



\subsection{First order transition due to breaking of the potential's spherical symmetry} 

\begin{figure}
\includegraphics[width=0.9\linewidth]{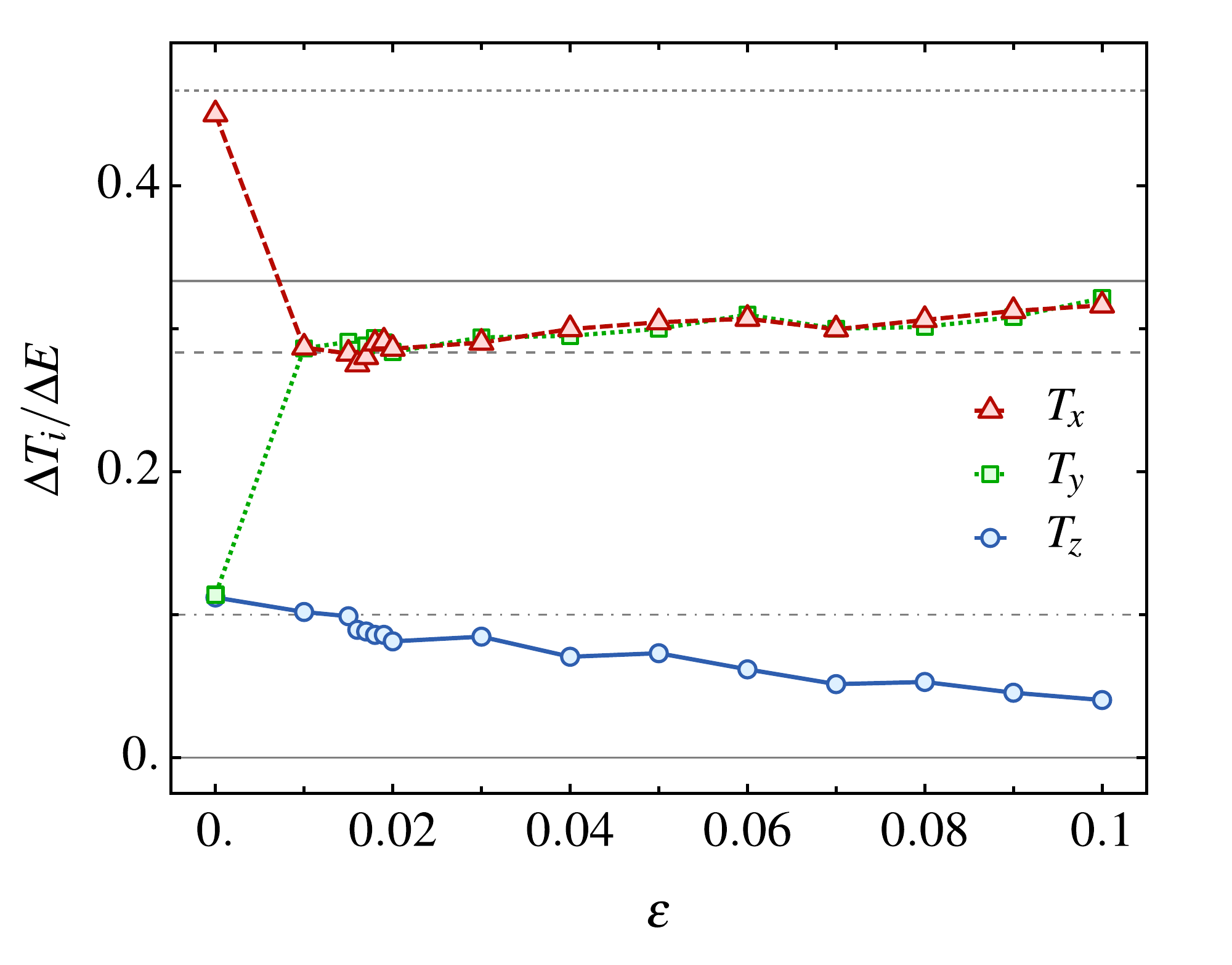}
\caption{Behavior of the final temperatures $\Delta T_i/\Delta E$ as a function of the anisotropy $\epsilon$ near the spherical limit after a kick along $x$. At $\epsilon=0$, $T_z=T_y$ after which there is a discontinuous change in the temperatures due to the breaking of spherical symmetry along $z$. Data are obtained by numerical simulation over 100 000 atoms. Dashed, dotted and plain lines correspond to the theoretical expectations of the fully isotropic (\ref{eq:sphericaltemperatures1}, \ref{eq:sphericaltemperatures2}), almost isotropic (\ref{eq:pz2_approx} and \ref{eq:finalsphericaltemperatures2}) and quadrupole geometries, respectively.
}
\label{fig:aniso}
\end{figure}

As we have seen, if we start with an isotropic equilibrium thermal distribution in a spherical trap ($\epsilon=0$) and we kick the gas along the $z$ direction then, when $t \rightarrow \infty$, we find that $T_x=T_y \neq T_z$. Likewise, by spherical symmetry, kicking along the $x$ direction will lead to the temperatures along the perpendicular directions being equal ($T_y=T_z \neq T_x$, see Fig.~\ref{fig:aniso}).

However, this is in seeming contradiction with the experimental results for the quadrupole case ($\epsilon=3$), see point 2. above and Fig.~\ref{fig:simul1}, where a kick along the $x$ direction leads to $T_x=T_y$. It seems that breaking the spherical symmetry by setting $\eps>0$ and making the $z$ direction unequal, enforces a cylindrical symmetry of the steady state gas distribution along the perpendicular directions after the kick. This discrepancy in behaviour indicates a discontinuous (first order) transition in gas behavior as a function of $\epsilon$ when going from spherical to non-spherical potentials.

To study this better we plot the three final temperatures after a kick along $x$ as a function of $\eps$ near $\eps=0$ (Fig.~\ref{fig:aniso}). We see that at $\eps=0$, $T_y=T_z < T_x$, as expected. However, for values of $\eps$ immediately above that, we find that $T_x=T_y>T_z$, the behavior of the quadrupole trap. In other words:
\be
\lim_{\eps \rightarrow 0} \lim_{t \rightarrow \infty} \langle p_x^2-p_y^2 \rangle_t   \neq  \lim_{t \rightarrow \infty} \lim_{\eps \rightarrow 0} \langle p_x^2-p_y^2 \rangle_t,
\ee
the lhs being zero and the rhs not. We will see that the reason for this is due to $\langle p_x^2-p_y^2 \rangle_t $ relaxing to zero with a  relaxation or dephasing time scale $\tau$ which diverges as $\eps \rightarrow 0$.


There is a characteristic relaxation time $\widetilde{\tau}$ before the momentum widths reach their final steady state value during which there is a gradual dephasing of the orbits of atoms with different angular momenta and energy in each plane. This timescale is related to the width of the thermal distribution and does not depend on $\eps$ as $\eps \rightarrow 0$. From dimensional analysis we see that $\widetilde{\tau} \sim \sqrt{T_0} \sim 1$.

However, there is a second much longer characteristic relaxation time $\tau$ during which $T_x$ and $T_y$ converge to each other and which was not present in the perfectly spherical case. This timescale appears because of the rotation (precession) of the orbital planes of each atom around the $z$ axis and is due to the potential's anisotropy. This phenomenon is known in astronomy when studying the orbit of satellites around slightly non-spherical planets, where it is called nodal precession \cite{Goldstein2002}.

\begin{figure}
\includegraphics[width=0.9\linewidth]{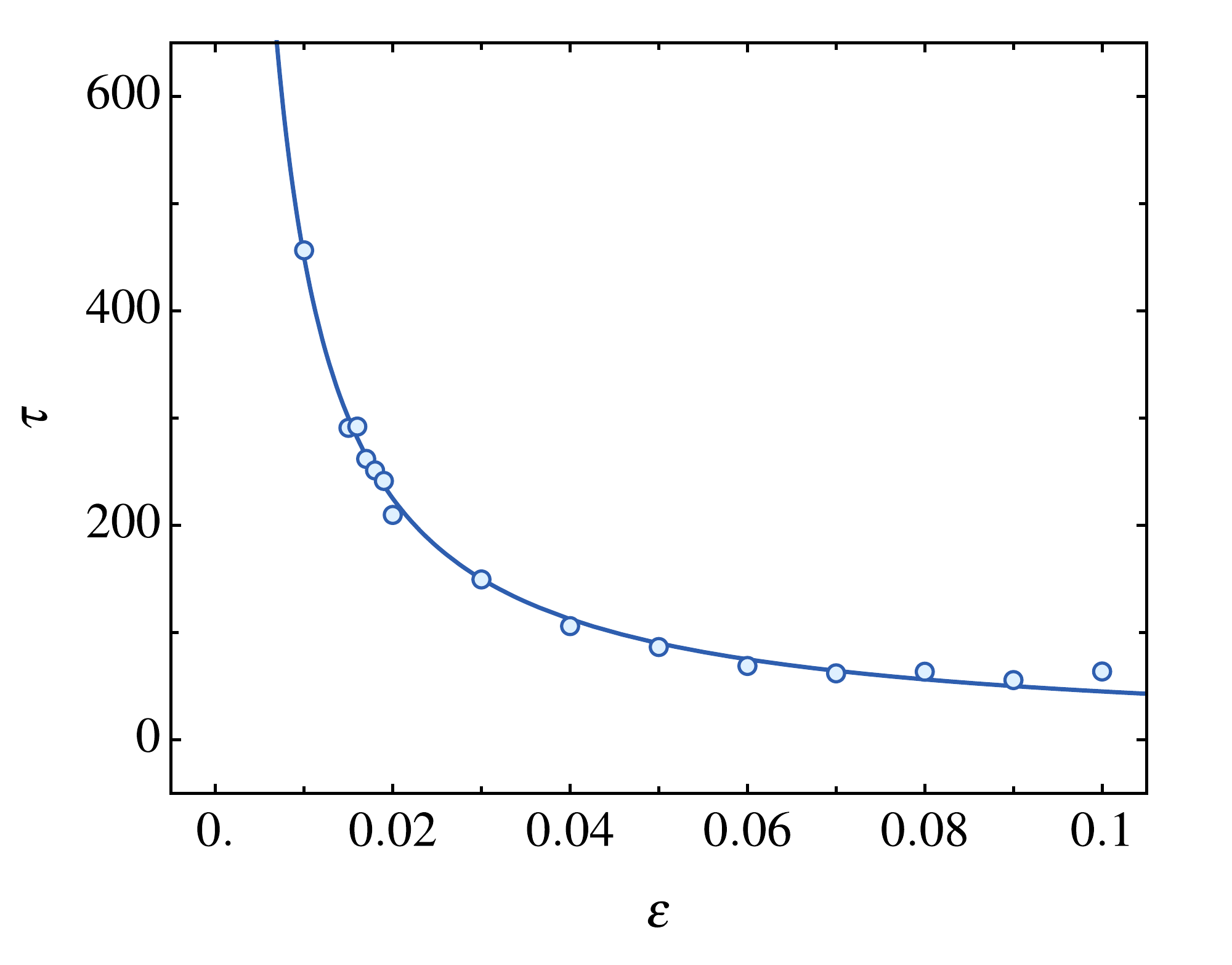}
\caption{Equilibration time as a function of the trap anisotropy $\epsilon$. The equilibration time is defined as the time required for $T_y$ to reach 99\% of the steady $T_x$ value. The solid line is a fit $A\times \eps^{-1}$ following the expression (\ref{eq:tau}). The best fit leads to an R-squared value of 0.99.}
\label{fig:time}
\end{figure}

For sufficiently small $\eps$ and at long times $t \gg \tau$, we expect that
$\langle p_x^2-p_y^2 \rangle_t $ will decay at long times as some function of $ t/\tau $, where the decay time scale is given by
\be
\tau \sim  \eps^\nu \label{eq:tau}.
\ee
The value of $\nu$ is independent of the kick strength if it is weak enough, and the dependence on $\sqrt{T_0}$ sets the dimensions of $\tau$. We show in Appendix \ref{sec:dephasing} that $\nu=-1$ so that $\tau \sim 1/\epsilon$; this is confirmed in Fig.~\ref{fig:time}.

While Bertrand's equilibrium leads to a higher temperature along the kicked direction, the orbital precession redistributes the energy equally between the $x$- and $y$-axes, leading eventually to the equilibration of $T_x$ and $T_y$.  The first process takes place in about 40 time units, while the latter process is much slower as the anisotropy is smaller, as shown in Fig.~\ref{fig:time}.

This analysis leads to quantitative predictions for the final temperatures at small anisotropy, particular for the temperature discontinuities. For $\epsilon \gtrsim 0$, the imparted energy gets first redistributed in the plane, before the orbital precession slowly equilibrates temperatures so that we can express the final temperatures in terms of the spherical temperatures \eqref{eq:sphericaltemperatures1}, \eqref{eq:sphericaltemperatures2}:
\begin{align}
 \Delta T _x ^{\epsilon \rightarrow 0} = \Delta T _y ^{\epsilon  \rightarrow 0} &= \frac{1}{2}\left( \Delta T _x ^{\epsilon = 0} + \Delta T _y ^{\epsilon = 0} \right) \\
 			&=  \frac{17}{60} \Delta E \quad \quad {\rm and} \\
 \Delta T _z ^{\epsilon  \rightarrow 0} & = \Delta T _z ^{\epsilon =0}\\
 		&	= \frac{1}{10} \Delta E. \label{eq:finalsphericaltemperatures2}
\end{align}
From here we can extract the matrix elements of \eqref{eq:theta_matrix_axial} since in the above equations $\Delta E=q^2_x/2$: $\theta_1=17/60$, $\theta_2=1/10$, which means that $\theta_3=7/15$.

This prediction is in very good agreement with the results presented in Fig.~\ref{fig:aniso} and is valid near $\epsilon=0$ as long as $\tau \gg \widetilde{\tau}$.
\section{Pancake limit} \label{sec:pancake}
In the previous section we analyzed the spherically symmetric case, which could be solved analytically. There is another case where the motion can be solved analytically, namely the limit when the confinement along the $z$-direction is strong $(\epsilon \gg 1)$. As we will see later, this case exhibits behavior which is much closer to the quadrupole.

We consider the case of strong confinement of the gas along the $z$ direction of the potential \eqref{eq:general_potential_with_epsilon} with $\epsilon \gg 1$ so that
\be
V(\br) = \sqrt{x^2 + y^2 + (1+\epsilon) z^2} \simeq \sqrt{\rho^2+\epsilon z^2},
\label{eq:pancakepot}
\ee
where we have used cylindrical coordinate $\rho=\sqrt{x^2+y^2}$. Since the potential is tightly confined, motion along the $z$ direction is fast compared to that in the plane. To check this we compare $t_z$, the period of oscillation along $z$, with the period of oscillation along $\rho$, $t_\rho$, the characteristic timescales of the two motions.

An atom whose motion is along the $x$-axis experiences a potential $V=x$, whereas if the atom moves along the $z$-axis, it sees a potential $V=\sqrt{\epsilon} z$. Assuming that both of these atoms have the same overall thermal energy $T$, then, in the first case, its period of oscillation is $ \propto \sqrt{E}$ whereas in the second case it is $ \propto \sqrt{E/\epsilon}$, so that the ratio of the two periods is $\epsilon \gg 1$ and therefore the approximation of considering the motion along $z$ to be fast compared with that in the plane is justified.

We start by analyzing the motion of a single atom. We split up the energy as
\be
E=\frac{p^2_\rho}{2}+\frac{p_\phi^2}{2\rho^2}+E_z, \label{eq:totalenergy}
\ee
where $p_\rho$ and $p_\phi$ are the canonically conjugate momenta and $\phi=\arctan(y/x)$ is the angle with the $x$-axis in the $xy$ plane. $E$ and $p_\phi$ are constants of the motion (the latter being the angular momentum $L_z$ which is always conserved due to the potential being independent of $\phi$). Also,
\be
E_z=\frac{p^2_z}{2}+\sqrt{\rho^2+\epsilon z^2}. \label{eq:energyz}
\ee
We now replace $p_z,z$ with the action-angle variables $I$, $\eta$ in the usual way. In particular
\begin{eqnarray}
I&=&\oint p_z \frac{dz}{2 \pi}= 4\int_0^{z_{\rm{max}}} p_z \frac{dz}{2 \pi} \nonumber \\
&=& \frac{2 \sqrt{2}}{\pi} \int^{\sqrt{\frac{E_z^2-\rho^2}{\eps}}}_0 \sqrt{E_z-\sqrt{\rho^2+\eps z^2}} dz \nonumber \\
& = & \frac{2 \sqrt{2}}{\pi \sqrt{\eps}} I_0 \label{eq:I}
\end{eqnarray}
with the definition
\begin{equation}
I_0 \equiv \int^{\sqrt{E_z^2-\rho^2}}_0 \sqrt{E_z-\sqrt{\rho^2+z^{\prime 2}}} dz^\prime \label{eq:I_0}
\end{equation}
where we made the substitution $z^\prime=\sqrt{\eps} z$ to show that $I \propto 1/\sqrt{\eps}$, since $I_0$ does not depend on $\eps$. Note that, for the same reason, in \eqref{eq:I_0} $E_z$ depends implicitly only on $I_0$ and $\rho$ but not on $\eps$.

The trajectory $p_z(z)$ is determined by \eqref{eq:energyz} and therefore depends on $E_z$ and $\rho$. Also, since $\rho$ is slowly varying, by the standard arguments, $I$ (or $I_0$) can be considered an adiabatic invariant (i.e.\ another constant) for the motion in the plane.

A simple approximate solution to \eqref{eq:I_0} which allows us to express $E_z$ explicitly in terms of $I_0$ and $\rho$ is
\beq
E_z(I_0,\rho)=\left(\frac{3}{2} I_0 +\rho^{3/2} \right)^{2/3}, \label{eq:I2}
\eeq
which allows us to rewrite \eqref{eq:totalenergy} as
\beq
E=\frac{p_\rho^2}{2}+\frac{p_\phi^2}{2\rho^2}+\left(\frac{3}{2}I_0 +\rho^{3/2} \right)^{2/3} \label{eq:totalenergy2}
\eeq
and we see that the effective potential for the radial motion is a sum of the centrifugal term plus a confining term increasing linearly at large distances.

Since we had originally three degrees of freedom, a particular trajectory is completely determined by the three integrals of the motion $E$, $p_\phi$ and $I_0$. Therefore, the time averaged in-plane kinetic energy
\be
\overline{ \frac{p^2_\rho}{2}+\frac{p_\phi^2}{2\rho^2}} \label{eq:kinetic_energy_in_plane}
\ee
is also determined by these constants.

It is now clear that the averaged kinetic energy is only a function of the constants of the motion $E$, $p_\phi$, and $I_0$ for that orbit.

The adiabatic principle tells us that the atom executes a motion in the plane under the effective potential $E_z$ given by \eqref{eq:I2}. Since $I_0$ is not the same for all atoms, each atom moves in slightly different potentials, labelled by their value of $I_0$.

When we apply a kick to an atom along $z$ at a time $t_0$, its in-plane momenta $p_\phi$, $p_\rho$ and its position $\rho$, $z$ are unchanged. What changes instead is its momentum $p_z$ and therefore its corresponding
kinetic energy $p_z^2/2 \rightarrow (p_z+q)^2/2=p_z^2/2 + p_z q + q^2/2$. After averaging over the whole gas, the term $p_z q$ drops out so that only the third term remains, an average increase of kinetic energy of $q^2/2$ per atom (and so of $E_z(\rho_0)$ as we see from \eqref{eq:energyz}). This has two effects: it changes the effective potential \eqref{eq:I2} and it increases the total energy $E$. Since $I_0$ is an increasing function of $E_z$, an increase of the kinetic energy along $z$ at $t_0$ implies an instantaneous change $I_0 \rightarrow I_0 + \Delta I_0$, $\Delta I_0>0$. In the subsequent motion, the effective potential is changed $E_z(I_0,\rho) \rightarrow E_z(I_0+\Delta I_0,\rho)$, transforming it into a shallower effective 2D potential as we can see from \eqref{eq:I2}. On the other hand, the increased kinetic energy also means an increased total energy $E \rightarrow E+\Delta E$:
\beq
\Delta E=E_z(I_0+ \Delta I_0,\rho_0)-E_z(I_0,\rho_0).
\eeq
The first effect leads naturally to a reduction in speed in the plane, i.e.\ a reduction in the average in-plane kinetic energy \eqref{eq:kinetic_energy_in_plane}. However, the increase $\Delta E$ has the opposite effect, that of increasing the average kinetic energy. This latter effect is dominant for atoms which were near the bottom of the potential at the moment of the kick, whereas the reduction in $\Delta E$ is most felt by those which were away from the center.

To determine what happens to the gas as a whole, we use numerical simulations. We compare the results of the pancake case after a kick along $z$ with a very large $\epsilon$ with the case of a 2D gas in the effective potential \eqref{eq:I2}, with the same number of atoms, temperature, and kick momentum $q$.


\begin{figure}
\centering
	\begin{overpic}[width=0.9\linewidth]{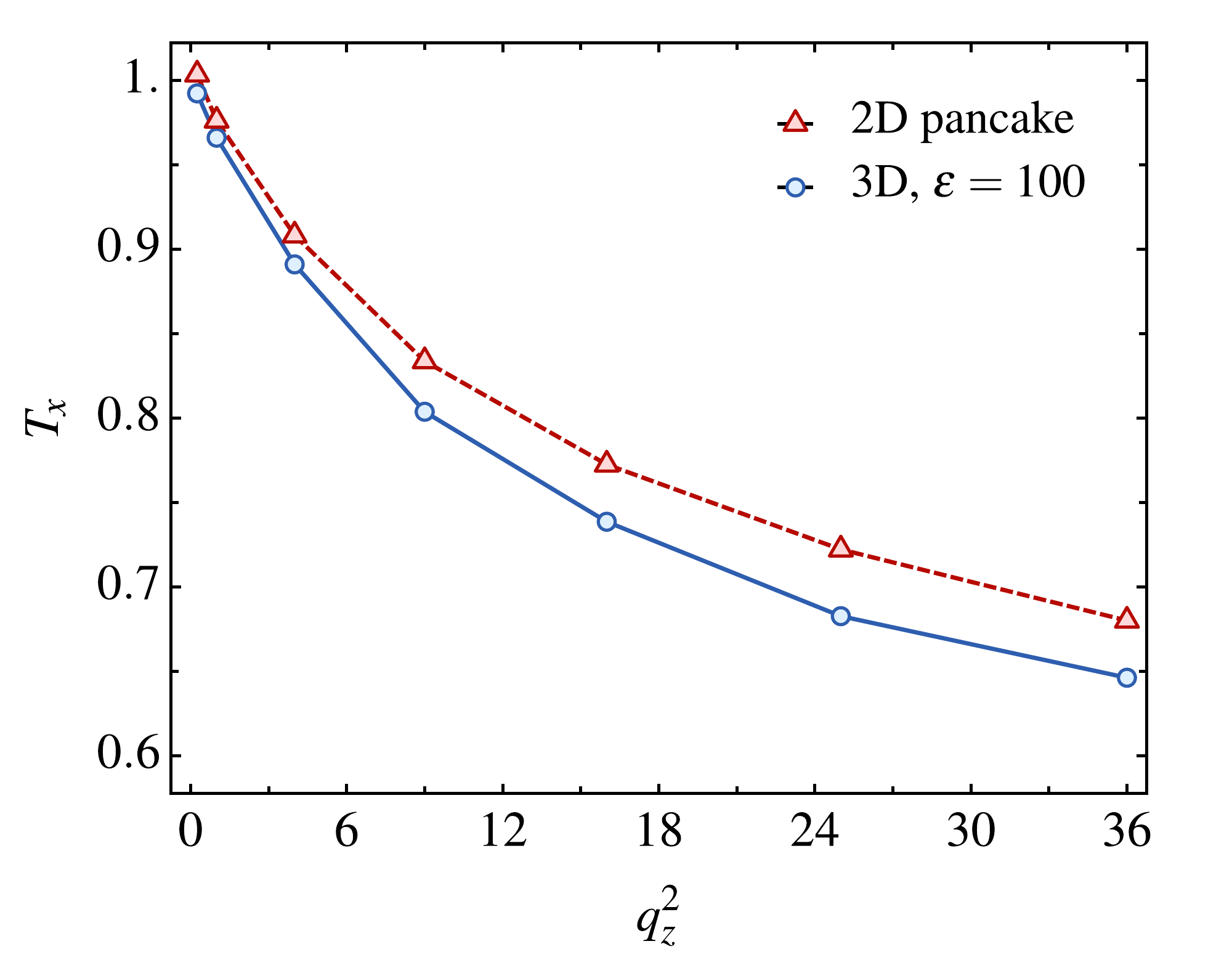}
		\put(1,73){\rm\bfseries a)}
	\end{overpic}
	
	\begin{overpic}[width=0.9\linewidth]{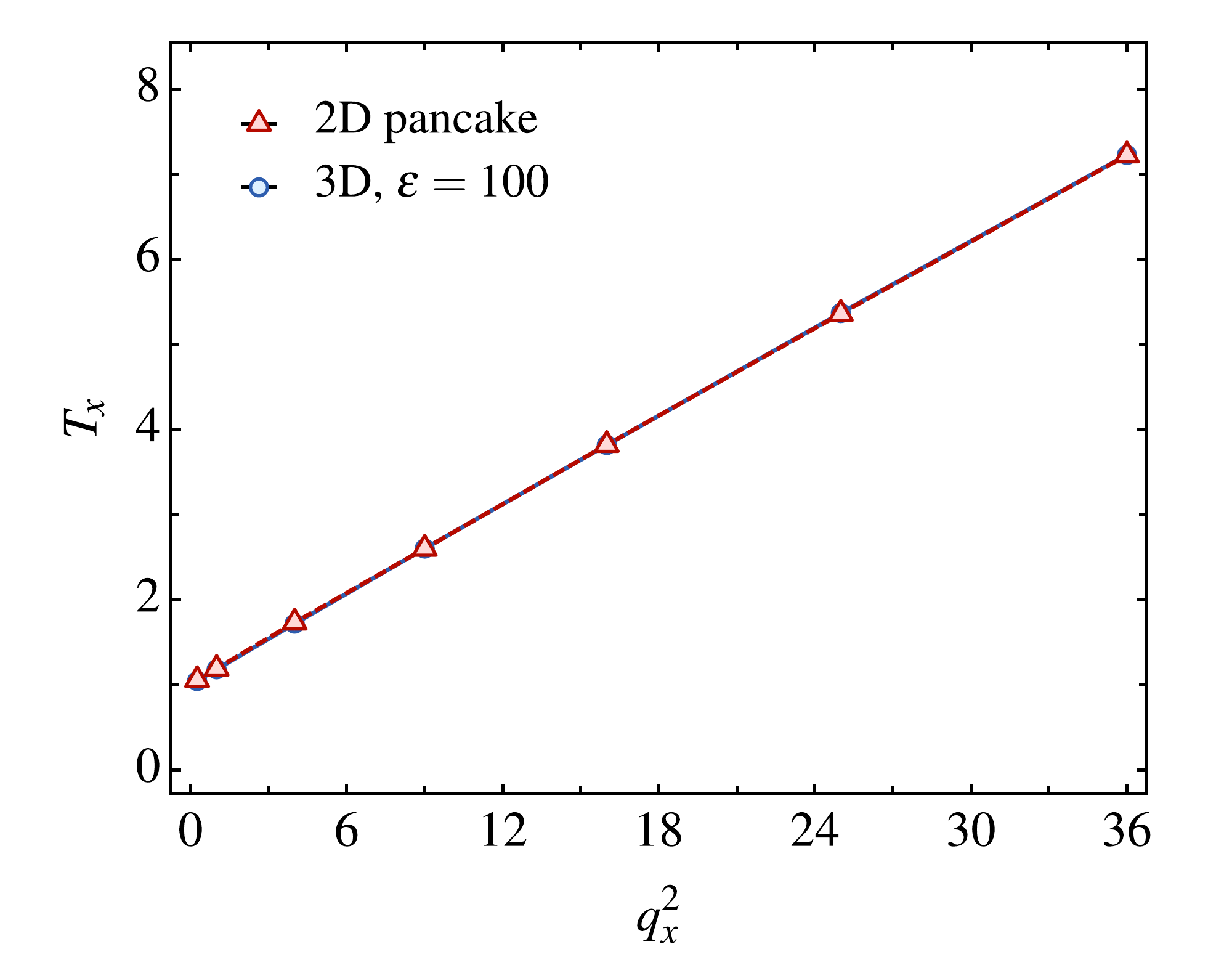}
		\put(1,73){\rm\bfseries b)}
	\end{overpic}
\caption{Comparison of the 2D potential given by \eqref{eq:I2} and the pancake potential.
Here $\epsilon=100$, but larger $\epsilon$ values produced consistent results. The plots are for $\langle p_x^2\rangle$ as a function of $q_z^2$ (top panel), and as a function of $q_x^2$ (bottom panel). Solid and dashed lines represent guides to the eye. Note that the two potentials give almost identical results for the heating along the kick direction and a small discrepancy for the orthogonal cooling.} \label{fig:pancake_2D}
\end{figure}

For the simulation of the 2D gas, we use the same initialization of the system as for the regular pancake case, namely the kicked Boltzmann distribution with potential \eqref{eq:pancakepot}. Then for each individual atom we evaluate $I_0$ via Eq.~\eqref{eq:I2}, where $E_z$ and $\rho$ are determined by the initial position and momentum coordinates of the atom. The subsequent time-evolution of each atom is governed by Eq.~\eqref{eq:totalenergy2}, where the last term corresponds to the new effective potential \eqref{eq:I2} which replaces the regular pancake potential \eqref{eq:pancakepot} ($I_0$ is assumed constant for each atom during the time-evolution). Note that we only evaluate the movement of the gas in the $xy$ plane in this approximation---the position and momentum coordinates in $z$-direction do not appear in the equations.


We are also able to use the same method to study the change in average kinetic energy along $z$ due to a kick in the plane along $x$.

Our findings are in Fig.~\ref{fig:pancake_2D}. We see that there is excellent agreement between the 3D pancake gas and the 2D case, especially for the heating along the direction of the kick. Although both show cooling of the transverse directions, the agreement is less good there, a fact which we attribute to the inexact ansatz \eqref{eq:I2}.

So the physical interpretation of the pancake case is clear: there is a slight overall cooling of the transverse directions when the gas is kicked along the $z$ direction due to the effective potential becoming shallower for the atoms closer to the center of the trap. This effect dominates over the heating of the atoms near the edges of the gas, although not by much so that the overall cooling is very small.

\section{Comparison of the potentials}

\begin{figure}

	\begin{overpic}[width=0.9\linewidth]{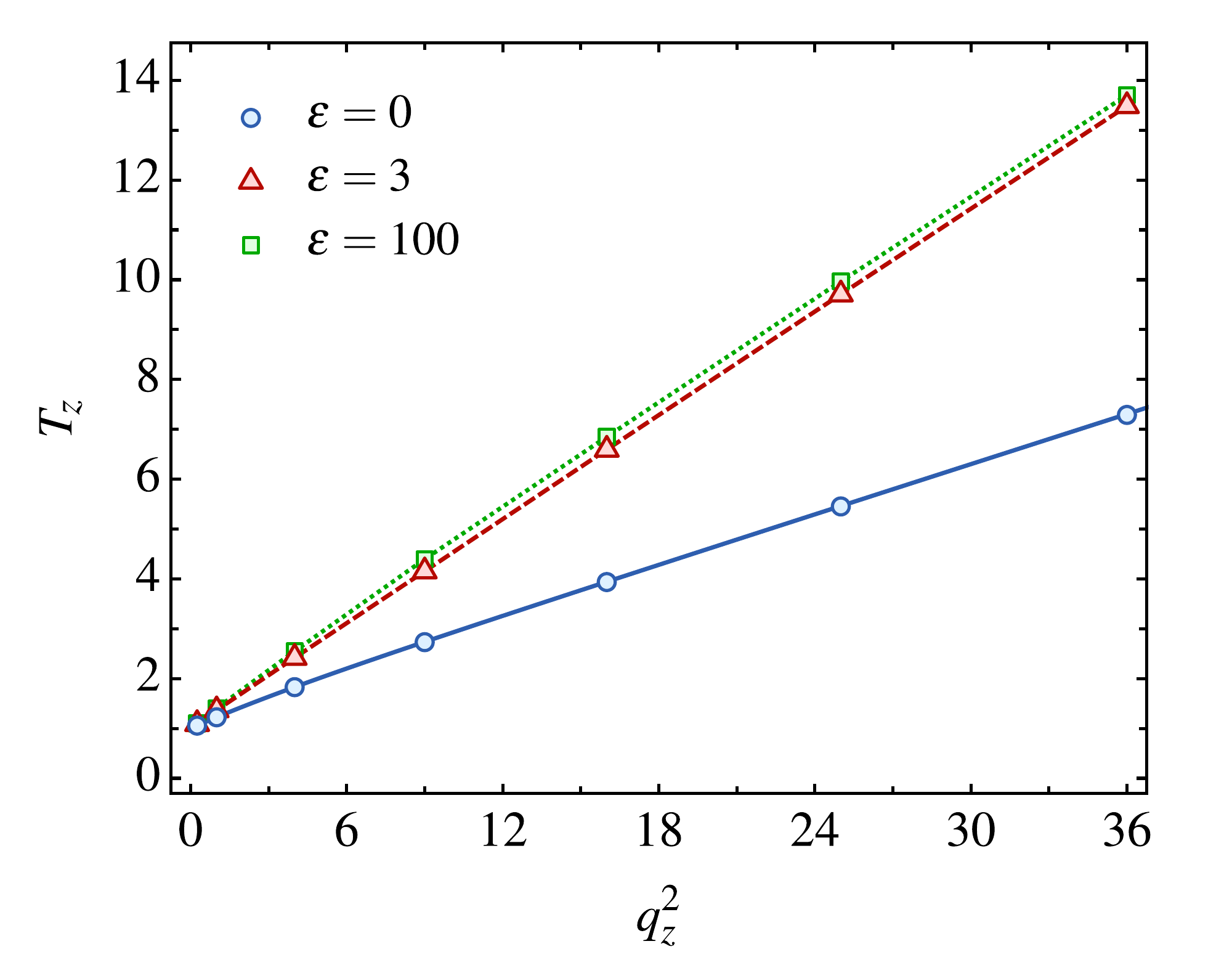}
		\put(1,73){\rm\bfseries a)}
	\end{overpic}
	
	\begin{overpic}[width=0.9\linewidth]{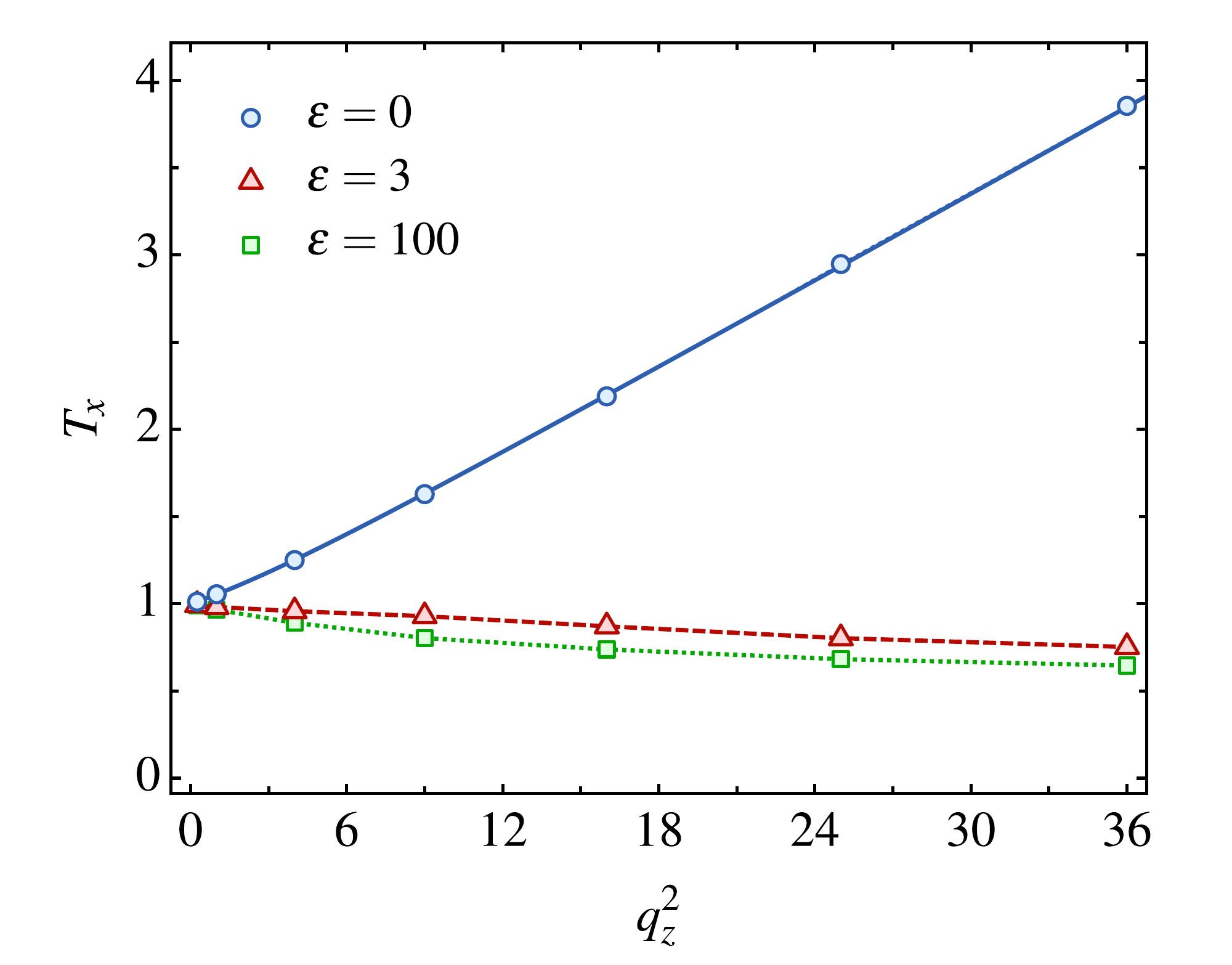}
		\put(1,73){\rm\bfseries b)}
	\end{overpic}

\caption{\label{fig:comparing_potentials} Comparison of the three types of potentials: spherical ($\epsilon=0$), quadrupole ($\epsilon=3$) and pancake ($\epsilon=100$) showing the much better agreement between the quadrupole and the pancake compared with the spherical case. The top panel shows $\langle p_z^2 \rangle$ and the bottom panel $\langle p_x^2 \rangle$ as a function of the momentum kick $q_z^2$ in $z$-direction. The blue solid lines are the analytical predictions (\ref{eq:average_pxy2_analytical}) and (\ref{eq:average_pz2_analytical}) for the spherical potential. Dashed and dotted lines for the other potentials are guides to the eye.}
\end{figure}

In Fig.~\ref{fig:comparing_potentials} we compare the quadrupole potential with the two others we have discussed, the spherical and the pancake. It is clear that the quadrupole behavior is much closer to that of the pancake so that, in this respect, it seems that the $\epsilon=3$ is already very close to the limit of $\epsilon=\infty$. There is a remarkably good quantitative agreement between the two cases. For example we obtain approximately the same heating and cooling in both the kicked and transverse directions. We find for the pancake $\Theta_{xx}=0.38$ and $\Theta_{zx}=-0.09$ which can be compared with the very similar values for the quadrupole $\Theta_{xx}=0.36$ and $\Theta_{zx}=-0.05$ mentioned in Sec.~\ref{sec:simulations}.

In Sec.~\ref{sec:simulations} we mentioned two puzzles: one was the anisotropy of the temperatures along the kicked and orthogonal directions in the quadrupole potential. Both the spherical and pancake potentials exhibit this. For the spherical case this is due to a geometric reason, the fact that the kick redistributes the atoms into planes which are more aligned with the direction of the kick. They will subsequently remain there due to the conservation of angular momentum. In the pancake case this is due to strong potential anisotropy coming from the large value of $\epsilon$. This latter reason is responsible also for the anisotropy in the quadrupole potential.

Also, in the spherical case we saw that the temperatures of the kicked direction and of the plane orthogonal to it were different. In the quadrupole case we find generally in all simulations that $T_x=T_y \neq T_z$. This was interpreted in terms of the 2D effective description where Bertrand's theorem applies; it leads naturally to the isotropy of the distribution in the $xy$ plane. Clearly the quadrupole gas has this behavior for the same reason.

The second puzzle was the apparent (near) separability of the kicked and orthogonal directions, i.e.\ that the kick energy is {\em not} redistributed into the momentum distributions of all directions but rather it is concentrated only in the kick direction. As we see from Fig.~\ref{fig:comparing_potentials}, the spherical and pancake potentials behave very differently: the pancake reproduces closely the quadrupole's separability (in fact a slight cooling of the orthogonal directions) while the spherical potential shows a clear heating of those directions. The reason for the separability in the quadrupole case can be traced to the 2D model where it is due to a near cancellation of the contributions of the atoms which, at the moment of the kick, are close to the center of the trap and are cooled and that of the atoms at the periphery which are heated.

\section{Conclusion}
We began this analysis with some puzzling experimental results for a non-interacting classical gas in a quadrupole trap whereby momentum kicks along one spatial direction were found to mostly affect only that direction, despite the fact that the potential is non-separable. By analyzing the extreme case of the spherical potential ($\epsilon=0$) we understood that, in 3D, the constants of the motion (e.g.\ the angular momentum components) can allow the system to retain a memory of the direction of the kick. Consequently, the long time momentum distribution can remain anisotropic in this {\em isotropic} system. However, this effect strongly depends on the dimensionality of the problem, and the situation is completely different in 2D, where Bertrand's theorem leads to an isotropic distribution. Furthermore, as soon as the potential becomes slightly anisotropic ($0<\epsilon \ll 1$), the competition between the in-plane isotropic behavior and the symmetry-induced precession of orbital planes results in a qualitatively different steady-state, which we were able to characterize analytically. Finally, we investigated the pancake limit (large $\epsilon$), which was shown numerically to be much closer to the experimental situation (i.e.\ the quadrupole potential). We were able to explain analytically, based on an effective potential, most of its characteristic features, including the peculiar cooling effect experienced by the transverse degrees of freedom with respect to the kick direction.

For future analysis we would like to investigate the apparent ``thermalization" of the gas after the kick as discussed in section \ref{sec:thermalisation} as well as studying more in detail the region between the spherical to pancake limits.


\bibliography{Bib/ECS,Bib/library,Bib/referencesOlga,Bib/ReferenceFred}

\begin{appendix}

\section{Coordinate transformation for the spherical case}
\label{sec:coordinates}

In a spherical potential the motion of each atom is confined to a plane through the origin and perpendicular to its angular momentum. To treat the gas in each plane as a separate system it is convenient to choose coordinates where the motion in each plane is described by in-plane 2D coordinates along orthogonal axes labelled $(u,v)$ with corresponding momenta $(p_u,p_v)$. To relate these to the rectangular coordinates we define two angles $\theta$ and $\phi$. $\theta$ is the angle between the $z$ axis and the $v$ axis. $\phi$ is the angle between the $x$-axis and the projection of the $v$-axis onto the $x$-$y$ plane. Both angles are in the interval [0,$\pi$]. The coordinate transformation is thus:
\begin{eqnarray} \label{eq:cartesian_plane_transformation}
 x & = & u \sin \phi + v \sin \theta \cos \phi \nonumber
\\
 y & = & - u \cos \phi + v \sin \theta \sin \phi \nonumber
\\
 z & = & v \cos \theta \nonumber
\\
 p_x & = & p_u \sin \phi + p_v \sin \theta \cos \phi \nonumber
\\
 p_y & = & - p_u \cos \phi + p_v \sin \theta \sin \phi \nonumber
\\
 p_z & = & p_v \cos \theta.
\end{eqnarray}
Note that it is not a canonical transformation since the Jacobian is:
\begin{equation} \label{eq:jacobian_cartesian_plane}
 J_1 = | (p_u v - p_v u) \cos \theta|.
\end{equation}
The $\cos \theta$ term has a simple interpretation: the angular density of planes having an angle $\theta$ with the $z$-axis is largest for small $\theta$ and drops to zero when $\theta=\pi/2$ since then there is only one plane perpendicular to the $z$-axis. In most cases, the calculation becomes simpler if we use polar coordinates in the plane:
\begin{eqnarray}
 u & = & r \cos \alpha_r \nonumber
\\
 v & = & r \sin \alpha_r \nonumber
\\
 p_u & = & p_r \cos \alpha_p \nonumber
\\
 p_v & = & p_r \sin \alpha_p \nonumber,
\end{eqnarray}
where $\alpha_r$ and $\alpha_p$ are in the interval (0,2$\pi$]. The transformation (\ref{eq:cartesian_plane_transformation}) becomes:
\begin{eqnarray} \label{eq:polar_plane_transformation}
 x & = & r \cos \alpha_r \sin \phi + r \sin \alpha_r \sin \theta \cos \phi \nonumber
\\
 y & = & - r \cos \alpha_r \cos \phi + r \sin \alpha_r \sin \theta \sin \phi \nonumber
\\
 z & = & r \sin \alpha_r \cos \theta \nonumber
\\
 p_x & = & p_r \cos \alpha_p \sin \phi + p_r \sin \alpha_p \sin \theta \cos \phi \nonumber
\\
 p_y & = & - p_r \cos \alpha_p \cos \phi + p_r \sin \alpha_p \sin \theta \sin \phi \nonumber
\\
 p_z & = & p_r \sin \alpha_p \cos \theta.
\end{eqnarray}
The Jacobian for the transformation (\ref{eq:polar_plane_transformation}) is
\begin{equation} \label{eq:jacobian_polar_plane}
 J_2 = r^2 p_r^2 |\sin (\alpha_r - \alpha_p) \cos \theta|.
\end{equation}
As the Boltzmann function $f(x,y,z,p_x,p_y,p_z,t)$ is normalized to unity, if we apply the transformations (\ref{eq:cartesian_plane_transformation}) or (\ref{eq:polar_plane_transformation}), the following quantities will also normalize to unity either in the $(u,v,p_u,p_v,\theta,\phi)$ or in the $(r,\alpha_r,p_r,\alpha_p,\theta,\phi)$ coordinates:
\begin{eqnarray}
 1 & = & \int_{0}^{\pi} d \phi \int_{0}^{\pi} d \theta \int_{-\infty}^{\infty} d p_u \int_{-\infty}^{\infty} d p_v \int_{-\infty}^{\infty} d u \int_{-\infty}^{\infty} d v J_1 f \nonumber
\\
 & = & \int_{0}^{\pi} d \phi \int_{0}^{\pi} d \theta \int_{0}^{2 \pi} d \alpha_p \int_{0}^{\infty} d p_r \int_{0}^{2 \pi} d \alpha_r \int_{0}^{\infty} d r J_2 f. \nonumber
\end{eqnarray}




The average energy $\langle E \rangle$ is given by
\begin{equation}
 \langle E \rangle (t) \equiv \langle E \rangle_{t} = \int d^3 \mathbf{r} \int d^3 \mathbf{p} f(\mathbf{r},\mathbf{p},t) E(\mathbf{r},\mathbf{p}).
\end{equation}
It is useful to define an energy $\langle E \rangle _{\rm plane}(\theta,\phi)$ which is the average energy of all the planes lying between $\theta$ and $\theta+d\theta$, $\phi$ and $\phi+d\phi$
\begin{equation}
 \langle E \rangle _{\rm plane} (\theta,\phi) \equiv \int_0^{2 \pi} d \alpha_r \int_0^{2 \pi} d \alpha_p \int_0^{\infty} dp_r \int_0^{\infty} dr J_2  f E 
\end{equation}
so that the total energy is, cf. (\ref{eq:energy_energy_plane_relationship}),
\begin{equation}
 \langle E \rangle = \int_0^{\pi} d \phi \int_0^{\pi} d \theta \langle E \rangle _{\rm plane}.
\end{equation}
Note that, after the kick, $\langle E \rangle _{\rm plane}$ is independent of time as the number of atoms on each plane is constant.


\section{Averages over momenta}
\label{sec:momenta}
Using the transformation (\ref{eq:cartesian_plane_transformation}), we can write the averages of $p_x^2$, $p_y^2$ and $p_z^2$ as:
\begin{widetext}
\begin{eqnarray}
\langle p_x^2 \rangle  &=&  \langle p_u^2 \sin^2 \phi \rangle+ \langle p_v^2 \sin^2 \theta \cos^2 \phi \rangle + 2 \langle p_u p_v \sin \theta \sin \phi \cos \phi \rangle,  \label{eq:average_px2}
\\
\langle p_y^2 \rangle  &=&  \langle p_u^2 \cos^2 \phi \rangle + \langle p_v^2 \sin^2 \theta \sin^2 \phi \rangle - 2 \langle p_u p_v \sin \theta \sin \phi \cos \phi \rangle,  \label{eq:average_py2}
\\
 \langle p_z^2 \rangle & = & \langle p_v^2 \cos^2 \theta \rangle \label{eq:average_pz2}.
\end{eqnarray}
The first term of (\ref{eq:average_px2}) can be written in terms of $\langle E \rangle _{\rm plane}$ using (\ref{eq:energy_plane_pu2_relation}),
\begin{eqnarray}
 \langle p_u^2 \sin^2 \phi \rangle &=&   \int_0^{\pi} d \phi \sin^2 \phi \int_0^{\pi} d \theta \int_{-\infty}^{\infty} d v\int_{-\infty}^{\infty} du \int_{-\infty}^{\infty} dp_v \int_{-\infty}^{\infty} dp_u J_1 f(u,v,p_u,p_v,\theta,\phi,t) p_u^2 \nonumber
\\
 & = & \int_0^{\pi} d \phi \sin^2 \phi \int_0^{\pi} d \theta \frac{m \alpha}{2 + \alpha} \langle E \rangle _{\rm plane}.
\end{eqnarray}
Using similar technique, (\ref{eq:average_px2}), (\ref{eq:average_py2}) and (\ref{eq:average_pz2}) can be written as:
\begin{eqnarray}
 \langle p_x^2 \rangle & = & \frac{m \alpha}{2 + \alpha} \int_0^{\pi} d \phi \sin^2 \phi \int_0^{\pi} d \theta \langle E \rangle _{\rm plane} + \frac{m \alpha}{2 + \alpha} \int_0^{\pi} d \phi \cos^2 \phi \int_0^{\pi} d \theta \sin^2 \theta \langle E \rangle _{\rm plane} + 2 \langle p_u p_v \sin \theta \sin \phi \cos \phi \rangle \label{eq:average_px2_2}
\\
 \langle p_y^2 \rangle & = & \frac{m \alpha}{2 + \alpha} \int_0^{\pi} d \phi \cos^2 \phi \int_0^{\pi} d \theta \langle E \rangle _{\rm plane}  + \frac{m \alpha}{2 + \alpha} \int_0^{\pi} d \phi \sin^2 \phi \int_0^{\pi} d \theta \sin^2 \theta \langle E \rangle _{\rm plane} - 2 \langle p_u p_v \sin \theta \sin \phi \cos \phi \rangle \label{eq:average_py2_2}
\\
 \langle p_z^2 \rangle & = & \frac{m \alpha}{2 + \alpha} \int_0^{\pi} d \phi \int_0^{\pi} d \theta \cos^2 \theta \langle E \rangle _{\rm plane}. \label{eq:average_pz2_2}
\end{eqnarray}
The third term in (\ref{eq:average_px2_2}) and (\ref{eq:average_py2_2}) when written explicitly is:
\begin{equation}
\langle p_u p_v \sin \theta \sin \phi \cos \phi \rangle = \int_0^{\pi} d \phi \sin \phi \cos \phi \int_0^{\pi} d \theta \sin \theta \int_{-\infty}^{\infty} dv \int_{-\infty}^{\infty} du \int_{-\infty}^{\infty} dp_v \int_{-\infty}^{\infty} dp_u J_1 f(u,v,p_u,p_v,\theta,\phi,t) p_u p_v. \label{eq:average_pu_pv}
\end{equation}
\end{widetext}
If $f(u,v,p_u,p_v,\theta,\phi,t)$ is independent of $\phi$ then (\ref{eq:average_pu_pv}) equals to zero. If $\langle E \rangle _{\rm plane}$ is also independent of $\phi$, (\ref{eq:average_px2_2}), (\ref{eq:average_py2_2}) and (\ref{eq:average_pz2_2}) will be reduced to:
\begin{eqnarray}
 \langle p_x^2 \rangle & = & \frac{m \alpha \pi}{2(2+\alpha)} \int_0^{\pi} d \theta \langle E \rangle _{\rm plane} (1 + \sin^2 \theta)  \nonumber
\\
 \langle p_y^2 \rangle & = & \frac{m \alpha \pi}{2(2+\alpha)} \int_0^{\pi} d \theta \langle E \rangle _{\rm plane} (1 + \sin^2 \theta)  \nonumber
\\
 \langle p_z^2 \rangle & = & \frac{m \alpha \pi}{2+\alpha} \int_0^{\pi} d \theta \langle E \rangle _{\rm plane} \cos^2 \theta.  \nonumber
\end{eqnarray}
We can see that $\langle p_x^2 \rangle$ and $\langle p_y^2 \rangle$ are equal. If we look at the ratio between $\langle p_x^2 \rangle$ and $\langle p_z^2 \rangle$:
\begin{eqnarray}
 \frac{\langle p_z^2 \rangle}{\langle p_x^2 \rangle} = 2 \frac{\int_0^{\pi} \langle E \rangle _{\rm plane} \cos^2 \theta}{\int_0^{\pi} \langle E \rangle _{\rm plane} (1 + \sin^2 \theta)},
\end{eqnarray}
as $\langle E \rangle _{\rm plane} (\theta) \geq 0$, the ratio of the integrals will be between 0 and 1, therefore we can derive an inequality:
\begin{equation} \label{eq:pz2_px2_ratio_constraint}
 \langle p_z^2 \rangle \leq 2 \langle p_x^2 \rangle.
\end{equation}

\section{Calculation of dephasing rate}
\label{sec:dephasing}

\subsection{Initial distribution}
We wish to study
\be
\langle p_x^2-p_y^2 \rangle_t=\int dr^3 dp^3 (p_x^2-p_y^2) f(\br,\bp,t).
\ee
However, since the gas is noninteracting, we can find time-dependent averages by following the position of individual atoms starting from an initial distribution of the gas and then averaging over that distribution. For example, to find $\langle p_x^2-p_y^2 \rangle_t$, instead of finding the time dependence of the Boltzmann distribution $f$, we calculate the quantity $p_x^2(t)- p_y^2(t)$ for each atom starting at the initial position $(\br_0,\bp_0)$ and then average over $\br_0,\bp_0$ weighted by the initial distribution:
\be
\langle p_x^2-p_y^2 \rangle_t=\int d\br_0 d\bp_0 \left[p_x^2(t)-p_y^2(t)\right]_{\br_0,\bp_0} \times  f(\br_0,\bp_0,t=0) \label{eq:average_0}
\ee

We take the initial distribution from \eqref{eq:xkickeddistribution} and expand in powers of $q$:
\begin{eqnarray}
f (t=0)&\propto& \exp \left(-\frac{(p_x-q)^2+p_y^2+p_z^2}{2T} \right) \exp \left(-\frac{V(x,y,z)}{T}\right) \nonumber \\
&=& \left(1+\frac{p_x q}{T} - \frac{q^2}{2T} + \frac{1}{2} \left( \frac{p_x q}{T} \right)^2 + O(q^3) \right) f_{q=0}
\end{eqnarray}
The first and third terms in the brackets do not contribute to \eqref{eq:average_0} since they are spherically symmetric and remain so during time evolution. The second term  $\propto p_x q $ is odd under the parity transformation $x,p_x \rightarrow -x,-p_x$. Since this parity is preserved under time evolution, the integral of this term is zero for all times. The only term that contributes to \eqref{eq:average_0} is the one proportional to $(p_x q)^2$. Therefore keeping the lowest nonzero term we obtain:
\be
f(t=0)= \frac{1}{2}\left(\frac{p_x q}{T} \right)^2f_{q=0}. \label{eq:initialf}
\ee

\subsection{Precession of the orbital planes}

To find this contribution we will make the crucial assumption that its orbital plane precesses slowly around the $z$-axis compared with the fast motion in each plane so that we are allowed to use the virial theorem to calculate averages in the plane as in subsection \ref{sec:Planes}.

An orbital plane which is precessing will be characterised by a constant angle $\theta$ and a rate of precession $\dot{\phi}$. To find this rate we consider a perturbation of the planar orbit in the limit of a small correction to the spherical potential. Since $|\epsilon| \ll 1$, we expand the potential in \eqref{eq:general_potential_with_epsilon} to order $\mathcal{O}(\epsilon)$:
\begin{equation} \label{eq:approx_potential}
 V = \sqrt{x^2 + y^2 + (1+\eps )z^2} \simeq r+ \Delta H
\end{equation}
with
\be
\Delta H \equiv \frac{z^2}{2 r} \epsilon
\ee
being the perturbation of the Hamiltonian.

The rate of rotation of the orbital plane $\dot{\phi}$ is given by \cite{Goldstein2002}:
\begin{equation} \label{eq:average_plane_rotation_rate_definition}
\dot{\phi} = \frac{1}{l} \frac{\partial \overline{\Delta H}}{\partial \cos i}
\end{equation}
where $i$ is the inclination of the orbital plane and is related to $\theta$ via $i = \frac{\pi}{2} - \theta$, $l$ is the magnitude of angular momentum. $\overline{\Delta H}$ is the time-averaged value of $\Delta H$ calculated using the orbits of the {\em unperturbed} Hamiltonian.

If the orbit in the plane were closed, the averaging would be over the period of the unperturbed orbit. In our potential, almost all orbits are open so the period is not well defined. However, if we average over a time on the order $\widetilde{\tau}$, then we can assume that the plane of the orbit (i.e. $\theta$) remains fixed during that time but that the time average over the motion in the plane has achieved a stationary value:
\be
\overline{\Delta H}  \equiv  \frac{1}{\widetilde{\tau}} \int_0^{\widetilde{\tau}}\Delta H dt
\ee
and applying the co-ordinate transformation in (\ref{eq:cartesian_plane_transformation}):
\begin{eqnarray}
 \overline{\frac{z^2}{\sqrt{x^2 + y^2 + z^2}} } & = & \overline{ \frac{v^2 \cos^2 \theta}{\sqrt{u^2 + v^2}} }
\\
 & = & \overline{ \frac{v^2}{\sqrt{u^2 + v^2}} } \cos^2 \theta.
\end{eqnarray}
The last step uses the fact that the angle of the plane $\theta$ has not changed appreciably after a time $t$.

Assuming that the averages over the time $\widetilde{\tau}$ are well reproduced using the virial theorem (since the unperturbed potential is simply $V_{\eps=0}=r$), we know that:
\begin{eqnarray}
 E & = & \frac{3}{2} \overline{V_{\eps=0}} \nonumber
\\
 & = & \frac{3}{2} \overline{ \sqrt{u^2 + v^2} } \nonumber
\\
 & = & \frac{3}{2}\left(  \overline{\frac{u^2}{\sqrt{u^2 + v^2}} } + \overline{ \frac{v^2}{\sqrt{u^2 + v^2}}} \right).
\end{eqnarray}

If we assume
\begin{equation}
\overline{\frac{u^2}{\sqrt{u^2 + v^2}}} =\overline{  \frac{v^2}{\sqrt{u^2 + v^2}} }
\end{equation}
we get
\begin{equation}
\overline{  \frac{v^2}{\sqrt{u^2 + v^2}}} = \frac{E}{3}.
\end{equation}

Therefore the perturbed Hamiltonian averaged over $\widetilde{\tau}$ is:
\begin{eqnarray}
 \overline{\Delta H} & = & \frac{\epsilon E}{6} \cos^2 \theta \nonumber
\\
 & = & \frac{\epsilon E}{6} \sin^2 i
\end{eqnarray}
and the rotation rate of the orbital plane is:
\begin{eqnarray}
\dot{\phi} & = & \frac{\epsilon E}{6 l} \frac{\partial}{\partial \cos i} (\sin^2 i) \nonumber
\\
 & = & - \frac{\epsilon E}{3 l} \cos i= - \frac{\epsilon E}{3 l} \sin \theta
\end{eqnarray}
so that the plane precesses at a constant rate.
For a single atom on a plane $\theta$, $\phi$ (similarly to the calculations in \ref{sec:momenta})
\beq
\langle p^2_x \rangle &=& \langle p^2_u \rangle \sin^2 \phi +  \langle p^2_v \rangle \sin^2 \theta \cos^2 \phi  \nonumber \\ &&+ 2 \langle p_u p_v \rangle \sin \theta \sin \phi \cos \theta
\eeq
and similarly for $\langle p^2_y \rangle $. The last term is zero $\langle p_u p_v \rangle \rightarrow 0$ because of the isotropy due to Bertrand's theorem. Using \eqref{eq:average_pxy2_final} with $\alpha=1$ we get
\be
\langle p_x^2-p_y^2 \rangle_t =\frac{E}{3} \cos^2\theta(\sin^2 \phi - \cos^2 \phi )
\ee
Here, $\theta$ is a constant whereas $\phi(t)=\phi_0+\dot{\phi} t$.

To find the total value, we use \eqref{eq:average_0}:
\beq
\langle p_x^2-p_y^2 \rangle_t&=&\int d\br_0 d\bp_0 (p_x^2(t)-p_y^2(t)) \times \left(\frac{p_{x0} q}{T} \right)^2f_{q=0} (\br_0,\bp_0)\nonumber \\
&=&\int d\br_0 d\bp_0  \frac{E(\br_0,\bp_0)}{3} \cos^2\theta_0(\sin^2 \phi (t) - \cos^2 \phi (t)) \nonumber \\
&&\times \left(\frac{p_{x0} q}{T} \right)^2 f_{q=0} (\br_0,\bp_0)
\eeq

We see that $\langle p_x^2-p_y^2 \rangle_t$ is a function of $\epsilon t$ so $\tau \sim 1/\eps$ since the only time dependence is through $\dot{\phi}$. So we conclude that in \eqref{eq:tau}, $\nu=-1$.

\end{appendix}

\end{document}